\documentclass[fleqn,usenatbib,useAMS]{mnras}

\usepackage{graphicx}	
\usepackage{amsmath}	
\usepackage{amssymb}	
\usepackage{multicol}        
\usepackage{bm}		
\usepackage{pdflscape}	
\usepackage{xcolor}
\usepackage{multirow}

\newcommand{\gx}{\textsc{Gadget-X}}

\newcommand{\Mbnd}{{\ifmmode{M_{\rm bnd}}\else{$M_{\rm bnd}$}\fi}}
\newcommand{\Mfof}{{\ifmmode{M_{\rm fof}}\else{$M_{\rm fof}$}\fi}}
\newcommand{\Mcrit}{{\ifmmode{M_{\rm 200c}}\else{$M_{\rm 200c}$}\fi}}
\newcommand{\Rcrit}{{\ifmmode{R_{\rm 200c}}\else{$R_{\rm 200c}$}\fi}}
\newcommand{\Rhost}{{\ifmmode{R_{\rm host}}\else{$R_{\rm host}$}\fi}}
\newcommand{\Mmean}{{\ifmmode{M_{\rm 200m}}\else{$M_{\rm 200m}$}\fi}}
\newcommand{\MBN}{{\ifmmode{M_{\rm BN98}}\else{$M_{\rm BN98}$}\fi}}

\newcommand{\hGpc}{{\ifmmode{h^{-1}{\rm Gpc}}\else{$h^{-1}$Gpc}\fi}}
\newcommand{\hMpc}{{\ifmmode{h^{-1}{\rm Mpc}}\else{$h^{-1}$Mpc}\fi}}
\newcommand{\hkpc}{{\ifmmode{h^{-1}{\rm kpc}}\else{$h^{-1}$kpc}\fi}}
\newcommand{\hMsun}{{\ifmmode{h^{-1}{\rm {M_{\odot}}}}\else{$h^{-1}{\rm{M_{\odot}}}$}\fi}}
\newcommand{\Mstar}{{\ifmmode{M_{*}}\else{$M_{*}$}\fi}}
\newcommand{\Mhalo}{{\ifmmode{M_{\rm Halo}}\else{$M_{\rm Halo}$}\fi}}
\newcommand{\Ngal}{{\ifmmode{N_{\rm gal}}\else{$N_{\rm gal}$}\fi}}
\newcommand{\Norph}{{\ifmmode{N_{\rm orphan}}\else{$N_{\rm orphan}$}\fi}}
\newcommand{\Nxorph}{{\ifmmode{N_{\rm non-orphan}}\else{$N_{\rm non-orphan}$}\fi}}
\newcommand{\Zsolar}{{\ifmmode{Z_{\odot}}\else{$Z_{\odot}$}\fi}}
\newcommand{\Msun}{{\ifmmode{{\rm {M_{\odot}}}}\else{${\rm{M_{\odot}}}$}\fi}}
\newcommand{\ltsima}{$\; \buildrel < \over \sim \;$}
\newcommand{\gtsima}{$\; \buildrel > \over \sim \;$}
\newcommand{\lsim}{\lower.5ex\hbox{\ltsima}}
\newcommand{\gsim}{\lower.5ex\hbox{\gtsima}}

\newcommand{\Tab}[1]{Table~\ref{#1}}
\newcommand{\Sec}[1]{Section~\ref{#1}}

\newcommand{\Eq}[1]{Eq.~(\ref{#1})}
\newcommand{\Fig}[1]{Fig.~\ref{#1}}
\newcommand{\beq}{\begin{equation}}
\newcommand{\eeq}{\end{equation}}

\usepackage[T1]{fontenc}
\usepackage{ae,aecompl}

\usepackage{newtxtext,newtxmath}

\title[Identifying bound galaxy pairs]{Galaxy pairs in The Three Hundred simulations II: studying bound ones and identifying them via machine learning}

\author[Contreras-Santos et al.]
{Ana Contreras-Santos,$^{1}$\thanks{Contact e-mail: \href{mailto:ana.contreras@uam.es}{ana.contreras@uam.es}}
Alexander Knebe,$^{1,2,3}$
Weiguang Cui,$^{1,4}$\thanks{Atracción de Talento fellow}
Roan Haggar,$^{5,6}$
\newauthor
Frazer Pearce,$^{5}$
Meghan Gray,$^{5}$
Marco De Petris,$^{7,8}$
and Gustavo Yepes$^{1,2}$
\\
$^{1}$Departamento de F\'isica Te\'{o}rica, M\'{o}dulo 15, Facultad de Ciencias, Universidad Aut\'{o}noma de Madrid, 28049 Madrid, Spain\\
$^{2}$Centro de Investigaci\'{o}n Avanzada en F\'isica Fundamental (CIAFF), Facultad de Ciencias, Universidad Aut\'{o}noma de Madrid, 28049 Madrid, Spain\\
$^{3}$International Centre for Radio Astronomy Research, University of Western Australia, 35 Stirling Highway, Crawley, Western Australia 6009, Australia\\
$^{4}$Institute for Astronomy, University of Edinburgh, Royal Observatory, Edinburgh EH9 3HJ, United Kingdom\\
$^{5}$School of Physics \& Astronomy, University of Nottingham, Nottingham NG7 2RD, United Kingdom\\
$^{6}$Waterloo Centre for Astrophysics, University of Waterloo, Waterloo, Ontario N2L 3G1, Canada\\
$^{7}$Dipartimento di Fisica, Sapienza Università di Roma, Piazzale Aldo Moro 5, 00185 Roma, Italy\\
$^{8}$I.N.A.F. - Osservatorio Astronomico di Roma, Via Frascati 33, 00040 Monteporzio Catone, Roma, Italy\\
}

\date{Last updated 2015 May 22; in original form 2013 September 5}

\pubyear{2021}

\begin{document}
\label{firstpage}
\pagerange{\pageref{firstpage}--\pageref{lastpage}}
\maketitle

\begin{abstract}
Using the data set of \textsc{The Three Hundred} project, i.e. 324 hydrodynamical resimulations of cluster-sized haloes and the regions of radius 15 $\hMpc$ around them, we study galaxy pairs in high-density environments. By projecting the galaxies' 3D coordinates onto a 2D plane, we apply observational techniques to find galaxy pairs. Based on a previous theoretical study on galaxy groups in the same simulations, we are able to classify the observed pairs into ``true'' or ``false'', depending on whether they are gravitationally bound or not. We find that the fraction of true pairs (purity) crucially depends on the specific thresholds used to find the pairs, ranging from around 30 to more than 80 per cent in the most restrictive case. Nevertheless, in these very restrictive cases, we see that the completeness of the sample is low, failing to find a significant number of true pairs. 
Therefore, we train a machine learning algorithm to help us to identify these true pairs based on the properties of the galaxies that constitute them. With the aid of the machine learning model trained with a set of properties of all the objects, we show that purity and completeness can be boosted significantly using the default observational thresholds. Furthermore, this machine learning model also reveals the properties that are most important when distinguishing true pairs, mainly the size and mass of the galaxies, their spin parameter, gas content and shape of their stellar components.
\end{abstract}




\section{Introduction}

Early studies showed that most observed galaxies can be classified into different types according to their morphology (mainly ellipticals or spirals, see Hubble's `tuning fork', \citealp{Hubble1936,Sandage1961}). However, not all of them fit this sequence perfectly. An early attempt to study these galaxies was the \textit{Atlas of Peculiar Galaxies} \citep{Arp1966}, which consists of images of more than 300 galaxies that show different peculiarities such as perturbations and deformations. Interactions and mergers between galaxies, which can affect them in different ways, have been shown to be the main force causing these peculiarities. Numerical simulations performed in the following years have helped to clarify this situation \citep{Toomre1972,BarnesHernquist1992} and to acknowledge the crucial role that these interactions play in galaxy formation and evolution (see e.g. \citealp{Conselice2014}, for a review). Today, the $\Lambda$ cold dark matter ($\Lambda$CDM) growth paradigm for the Universe describes a hierarchical model of structure formation, such that galaxies are the result of many mergers of smaller objects \citep{White1978,Frenk2012}.

In this context, identifying galaxy mergers in the sky is a fundamental task. This includes not only galaxies that have already merged, but also galaxies that will merge in the future, so that both pre- and post-merger phases can be investigated. From an observational perspective, a common way to identify merger candidates is using galaxies that are close to each other in the sky, which are generally strongly related to pre-merger stages. Hence, several efforts have been devoted to studying close pairs of galaxies, based on certain selection criteria, usually a maximum projected separation,
maximum velocity separation and minimum mass ratio  (e.g. \citealp{Carlberg1994,Patton2000,Kartaltepe2007}). Some studies try to additionally select pairs that are going to merge by applying some asymmetry conditions \citep{Lotz2008,Conselice2008}. Assuming a certain merger timescale, the merger rate of galaxies can be estimated from this selection of pairs \citep{Patton2005,DePropris2007,Lopez-Sanjuan2011,Lopez-Sanjuan2015,Stott2013,Casteels2014}. 
If no additional selection of the pairs is made, the merger rate can also be estimated by assuming a different timescale, which may not have a physical meaning (e.g. \citealp{Kitzbichler-White2008,Xu2012,Mundy2017,Duncan2019,Husko2022}).

It has also been shown that galaxy pairs can actually affect the physical properties of the involved galaxies. In general, galaxies with close companions exhibit enhanced star formation \citep{Barton2000,Li2008a,Scudder2012,Patton2013,Pan2018}, diluted metallicities \citep{Kewley2010,Rupke2010,Bustamante2020} and, in some cases, an enhancement of the AGN (active galactic nuclei) activity \citep{Silverman2011,Cotini2013,Ellison2019}. These tendencies are believed to be kept even when the galaxies are in high-density environments \citep{Perez2006a,Alonso2006,Alonso2012}.

As with any other observation, when examining close pairs in the sky we have to be aware that we are actually observing a 2D projection of the 3D physical situation. Thus there can be projection effects that, for instance, create spurious pairs that are not close in real space or not physically bound. A more theoretical approach to identifying merger candidates is by looking for galaxies that are gravitationally bound to each other. Although being bound does not guarantee that the galaxies will merge, as this also depends on other conditions concerning the galaxies’ orbit (see e.g. \citealp{Barnes1992}), it is a necessary condition that can thus be relevant to check. However, this kind of analysis requires a great deal of theoretical information. Hence, it is better performed with numerical simulations rather than observations of the sky. Cosmological simulations, where the galaxies evolve naturally in a given environment, allow for easy identification of bound galaxies, and analysis of their properties \citep{Aarseth1980,Moreno2013,Haggar2021}.

We have already devoted a paper \citep{Contreras2022b} to using numerical simulations to study this distinction between what we named `good' and `bad' pairs, i.e. galaxy pairs with a physical separation within an allowed 2D range, and those with higher 3D separations than allowed. In this previous paper, we studied how many of the galaxy pairs we observe in a cluster environment are close in the physical distance as well. We also analysed the properties of the pairs, and how they differed for `good' and `bad' pairs.

This procedure provided useful information regarding the identification of pairs of galaxies in the sky, and the properties to be expected of them. Nevertheless, it was still an identification based only on physical distance, with no information on the boundness mentioned earlier. As such, two galaxies can be close in distance but with no strong attachment, generally referred to as a flyby. These kinds of interactions have been shown to become more relevant as the Universe expands, and therefore at later times, when mergers become less frequent \citep{vandenBergh1996,Murali2002,Sinha2012}. Although flybys can also affect galaxy morphology and properties to varying degrees \citep{Berentzen2004,Lang2014,Duc2013,Choi2017}, it is important to distinguish them from mergers, where the two galaxies will end up as a single object. 
For this reason, in this work, we will go one step further from our previous work and identify galaxy pairs that are also gravitationally bound. For that we use the simulations provided by \textsc{The Three Hundred} project\footnote{\url{https://the300-project.org/}} (as we already described in \citealp{Contreras2022b}).

\textsc{The Three Hundred} simulations consist of a set of 324 hydrodynamical re-simulations of the most massive clusters in a dark-matter-only cosmological simulation. These clusters reside in a high-density environment where the interactions between galaxies are more frequent and can be especially important. Although on large scales clusters are dark matter -- and hence gravity -- dominated, on smaller scales the baryonic components also play an important role (see \citealp{Kravtsov2012}, for a review on galaxy clusters). This leads to several different phenomena that drive galaxy evolution, making clusters very interesting environments to study galaxy interactions \citep{Gnedin2003,Park2009,Boselli2014}.

Using \textsc{The Three Hundred} simulations, we are going to first identify galaxy pairs in the sky using observational techniques. Then we will use the full information from the simulations to classify them as bound or not. Finally, we will develop a method to improve the performance of the two previous steps. This approach to identifying galaxy pairs in the sky maximises the fraction of them that are actually bound, without missing a very significant amount of the physical pairs. In order to do this, we will train a machine learning model with the data from the simulations and see how it can be used to classify observational data.

Artificial intelligence is a field that has expanded and gained importance very rapidly over the last decade. Machine learning (ML) in particular is a very valuable tool that can be used to address many problems from data analysis to image recognition. As with any other new technologies, many of the developments are now commonly applied in astrophysics (see e.g. \citealp{FlukeJacobs2020} for a review). The general idea is to create models that are capable of learning complex relationships between input and output variables, and that can then be used to make predictions on unseen data. 
For classification problems, ML can help avoid visual inspections, which can be both time-consuming and non-objective. Different algorithms have been used to estimate galaxies' morphology \citep{Banerji2010,Huertas-Company2015,DominguezSanchez2018}, as well as identifying galaxy mergers \citep{Ackermann2018,Pearson2019a,Bottrell2019,Ciprijanovic2020} or AGN hosts \citep{Faisst2019,Chang2021}. 
In this context, simulations have the advantage that they allow for comparison against ground truth, and thus allow us to evaluate the performance of the model when trained on mock observations (see e.g. \citealp{Snyder2019} or \citealp{Rose2022}). 

In this paper, we combine observational techniques to identify galaxy pairs with ML techniques to classify them and evaluate the results. 
The content is organized as follows. In \Sec{sec:data}, we present the details of the simulation and the halo catalogues used to identify the haloes and their properties. In \Sec{sec:method} we present the method used to find close pairs of galaxies. For the pairs found this way, in \Sec{sec:stats} we compare them against the theoretical work by \citet{Haggar2021} and analyse the results. In \Sec{sec:results} we introduce the ML algorithm used to classify the pairs and describe how it is used and how it improves the previous performance. Finally, in \Sec{sec:conclusions}, we summarize and discuss our results.

\section{The Data} \label{sec:data}

\subsection{The Three Hundred Simulations}
The simulations used in this work are part of \textsc{The Three Hundred} project, which consists of a set of 324 theoretically modelled galaxy clusters and the regions around them. This data set was presented in an introductory paper by \citet{Cui18}, and several other papers have been published based on this data (see e.g. \citealp{Mostoghiu18,Kuchner2021,Contreras2022a,Cui2022}). Here we will summarise the main aspects of the simulations, but we refer the reader to these works for further details about \textsc{The Three Hundred} project.

The 324 clusters in \textsc{The Three Hundred} sample were based on the DM-only MDPL2 MultiDark Simulation\footnote{The MultiDark simulations – incl. the MDPL2 used here – are publicly available at \url{https://www.cosmosim.org}} \citep{Klypin16}, which is a periodic cube of comoving length 1 $\hGpc$ containing $3840^3$ DM particles, each of mass $1.5\times 10^9$ $\hMsun$. The Plummer equivalent softening of this simulation is 6.5 $\hkpc$ and its cosmological parameters are based on the Planck 2015 cosmology \citep{Planck2015}. The 324 objects with the largest halo virial mass\footnote{The halo virial mass is defined as the mass enclosed inside an overdensity of $\sim$98 times the critical density of the universe \citep{Bryan98}} at $z=0$ ($M_\mathrm{vir}\gtrapprox 8 \times 10^{14} \hMsun$) were selected from this simulation, together with spherical regions of radius 15 $\hMpc$ around them. Within these regions, the initial DM particles were traced back to their initial conditions and then split into dark matter and gas particles according to the cosmological baryon fraction. The resulting mass resolution for these particles is $m_\mathrm{DM}=1.27 \times 10^9$ $\hMsun$ for dark matter and $m_\mathrm{gas}=2.36 \times 10^8$ $\hMsun$ for gas particles. Moreover, to reduce the computational cost while keeping large-scale tidal effects, dark matter particles outside these regions were degraded to a lower resolution. Then, each cluster region was re-simulated from the initial conditions but including full hydrodynamics using the SPH (Smoothed particle hydrodynamics) code \gx. The output produced consists of 129 snapshots between $z=16.98$ and $z=0$ for each of the 324 regions. 

\gx, the code used for the re-simulations, is a modified version of the non-public \textsc{Gadget3} code \citep{Murante2010,Rasia2015,Planelles2017,Biffi2017}. This uses the \textsc{Gadget3} Tree-PM gravity solver (an advanced version of the \textsc{Gadget2} code; \citealp{springel_gadget2_2005}) to evolve dark matter as well as baryons. It also includes an improved SPH scheme with artificial thermal diffusion, time-dependent artificial viscosity, high-order Wendland C4 interpolating kernel and wake-up scheme (see \citealp{Beck2016} and \citealp{Sembolini2016} for a presentation of the performance of this SPH algorithm). Star formation follows the classical \citet{Springel03} prescription and is implemented in a stochastic way which leads to varying star particle masses of order $m_{*} \sim 4 \times 10^7$ $\hMsun$. Stellar evolution and metal enrichment is originally described in \citet{Tornatore2007}, with further updates described in \citet{Murante2010} and \citet{Rasia2015}. SNeII are the only contributor to kinetic stellar feedback, which follows the prescription of \citet{Springel03}, with a fixed wind velocity of 350 km/s. Black hole (BH) growth and AGN feedback are implemented following \citet{Steinborn2015}, where supermassive black holes (SMBHs) grow via Bondi-Hoyle-like gas accretion (Eddington limited), with the model distinguishing between a cold and a hot component.

\subsection{The Halo Catalogues}
To identify the haloes in \textsc{The Three Hundred} simulations we use the open-source halo finder AHF (Amiga Halo Finder, \citealp{Gill04a,Knollmann09}). AHF finds potential halo centres as local overdensities in an adaptively smoothed density field, and thus automatically identifies haloes and substructures (see \citealp{Knebe11} for more details on halo finders). The radius at overdensity 200, $R_{200}$ is defined as the radius $r$ at which the density $\rho(r)=M(<r)/(4\pi r^3/3)$ drops below 200 times the critical density of the Universe at a given redshift, $\rho_{\mathrm{crit}}$. This is computed for each (sub)halo found by AHF, together with the corresponding enclosed mass $M_{200}$, as well as the analogous quantities for an overdensity of 500. For the substructure, AHF defines subhaloes as haloes that lie within $R_{200}$ of a more massive halo, which is called the host halo. The mass of this host halo then includes the masses of all the subhaloes contained within it.
Apart from mass and radius, AHF also allows for other properties to be generated for each (sub)halo, considering their gas, stars and dark matter particles. Properties such as peculiar velocities or angular momentum are based on all the bound particles that account for a halo. 

To obtain additional information, the stellar population synthesis code \textsc{stardust} (see \citealp{Devriendt99}, and references therein) can be used to produce luminosities (and magnitudes) in any spectral band, by considering the contribution of all the individual stellar particles and assuming a Kennicutt initial mass function \citep{Kennicutt98}. We finally note that in this work we only use the simulation snapshots at $z=0$, so no merger trees are needed to trace the haloes across cosmic time. 

\subsection{\textsc{Caesar} catalogues}
Apart from the AHF halo catalogues, to include further properties of the galaxies, also the \textsc{Caesar} galaxy finder was run on \textsc{The Three Hundred} data set. \textsc{Caesar}\footnote{\url{https://github.com/dnarayanan/caesar}} is a \texttt{yt}-based  python package for analysing the outputs from cosmological simulations (\texttt{yt} is an open source, astrophysical analysis and visualization tool, cf. \citealp{Turk2011}). Originally, \textsc{Caesar} provides a halo catalogue generated using a 3D Friend of Friend (FoF) algorithm with the galaxy catalogue using a 6D (in both spatial and velocity fields) FoF. It takes as input a single snapshot from a simulation and outputs an HDF5 catalogue containing a list of galaxy and halo properties, including physical and photometric properties for each object. To be consistent with the AHF catalogue, we run \textsc{Caesar} to only identify galaxies by using the (sub)halo information from AHF. 
Thus, the galaxies from \textsc{Caesar} can be precisely matched to the subhalos from AHF with their IDs. This way we can combine AHF and \textsc{Caesar} properties for all the objects we work with. This kind of joint analysis with the two catalogues has already been done by \citet{Cui2022} in \textsc{The Three Hundred} simulations. 

One limitation of our study that needs to be mentioned regards the numerical resolution. In previous convergence studies it is shown that, although halo mass is very stable for halos down to 20-30 particles, other individual properties of halos exhibit a more significant scatter. \citet{Trenti2010} show that, for additional halo properties like core density, virial radius and angular momentum, $N \gtrapprox 100 - 400$ particles is needed to guarantee a scatter below 20 per cent and achieve convergence. Regarding the shape, \citet{Allgood2006} find that, for halos with $\sim 300$ particles, the error in estimating the shape can be around 10 per cent.
With the resolution of \textsc{The Three Hundred} simulations, the galaxies used in this study range between 100 and 1000 stellar particles. Therefore, although some of them have stellar properties below the suggested resolution limits, our sample includes a significant number of galaxies above these limits. Nevertheless, the results should be interpreted with caution, being aware of this non-negligible scatter due to low number of particles.

\section{Methodology} \label{sec:method}

In this Section, we first present the way observers find close pairs of galaxies in the sky. Then we explain how we apply this same method to \textsc{The Three Hundred} data set, and how we can compare and correlate it with the more theoretical information that can be extracted directly from simulations. 

\subsection{Finding pairs in observations}
From an observer's perspective, defining two galaxies as a close pair depends on two quantities: their projected separation in the sky and their separation in velocity along the line of sight. In general, some kind of selection criteria is first applied to the galaxies, for instance, based on their luminosity or stellar mass, and then the remaining galaxies are paired based on these two quantities. This way, two galaxies are defined as close if their projected separation and line-of-sight velocity separation are within certain values selected as thresholds, which we will designate as $r_\mathrm{sep}$ and $v_\mathrm{sep}$, respectively. In the literature, the specific values used for these thresholds depend strongly on the particulars of the study in question. The values in $r_\mathrm{sep}$ range from 20 $\hkpc$ in works about galaxy mergers themselves (e.g. \citealp{Robotham2014}) up to 2 Mpc when the focus is on the effects of interactions and companions (e.g. \citealp{Patton2016}). The range is also wide for $v_\mathrm{sep}$, with the values depending on how the redshifts of the galaxies are determined. When the redshifts are determined spectroscopically, they are accurate enough to apply cuts in km/s, that can range from 250 to 1000 km/s. However, studies where photometric redshifts are used prefer to apply cuts in redshift separation, such as $\Delta z < 0.2$ \citep{Williams2011}. In this case, the conversion to line-of-sight velocity can yield differences much larger than 1000 km/s. 

Finally, most works on galaxy mergers also include separating them into major or minor, based on the stellar mass ratio (sometimes luminosity or flux) of the two involved galaxies. In general, pairs where this ratio is below 1:10 are not major nor minor but instead are discarded. \citet{Husko2022} provide a useful summary of different observational close pair studies, indicating the different selection criteria applied in each of them.

\subsection{Application to simulations}
Having this observational method already defined, we now want to implement it in our simulations, replicating it as much as possible. 
In our previous work \citet{Contreras2022b} we already applied this procedure to find pairs and groups of close galaxies in \textsc{The Three Hundred} data set. The methodology followed here will be essentially the same but, since there are some slight differences, we will nevertheless describe the process here. We still refer the reader to that work for further information regarding the statistics of 2D pairs and groups found.

As in \citet{Contreras2022b}, we first apply a selection of the objects we are going to work with from the simulations. Throughout this work, we will use the word ``galaxy'' to refer to the objects in the hydrodynamical simulations, including both their stellar and dark matter components. For each of the 324 clusters in \textsc{The Three Hundred} data set, we select the galaxies that are within $5R_{200}$ of the main cluster centre, and we apply a stellar mass cut $M_\mathrm{*}>10^{9.5}$ $\Msun$, similarly to what is done in observational studies (see \citealp{Cui18}, for the stellar mass function of all the galaxies in our data set). We also remove from our selection all the objects in the simulations with $M_{200} > 10^{13}$ $\Msun$, so that we do not include the most massive objects like galaxy clusters, and work only with galaxy-galaxy pairs. 
Although the interaction of satellites with the brightest cluster galaxies (BCGs) can also be of great interest, for consistency we prefer to not allow for such pairs in this study, since BCGs have been shown to be different from typical elliptical galaxies, in both their formation and evolution mechanisms \citep{LinMohr2004,Brough2005}. 
These selections leave us with a total number of galaxies between $\sim 400$ and 1200 depending on the cluster, which is still enough to have a significant number of pairs.

In order to find pairs within the selected galaxies, we first create `mock observations' by projecting the galaxies' 3D coordinates into a 2D plane. For simplicity, we will always project into the XY plane. In \citet{Contreras2022b} we randomly rotated the coordinates before projecting them, so that we obtained 100 different random projections for each cluster. In this work, we only create one projection for each of the 324 clusters, since this provides a large enough sample size whilst also simplifying the process significantly. We do not expect any differences in our results due to the projection direction; we have checked that, although having fewer statistics leads to higher scatter, the main results hold when doing this. Our pair-finding algorithm is based on the two parameters $r_\mathrm{sep}$ and $v_\mathrm{sep}$. The spatial separation between two galaxies is simply their distance in the XY plane, while for the velocity separation we consider two contributions to the line-of-sight velocity: the peculiar velocity of the galaxies along the $z$-axis (given by AHF) and the difference in recession velocities due to the Hubble flow (computed as $H \cdot \textbf{r}$, $H$ being the Hubble constant and $\textbf{r}$ the coordinates of each object). Two galaxies are considered close if their distance and velocity separations are below $r_\mathrm{sep}$ and $v_\mathrm{sep}$ respectively. For further research into the importance of these parameters, we use three different thresholds for each, so that they can be combined in 9 different ways. For the 2D-spatial separation, we use the values 20, 50 and 100 $\hkpc$, while for the velocity separation we use 300, 500 and 1000 km/s. 

Although most observational studies also apply a cut in the stellar mass ratio of the pairs, we prefer not to apply a similar cut and work with all the pairs found. This way we do not bias our sample towards pairs with similar masses, but rather keep the mass ratio as another feature to describe the galaxy pair, whose relevance can be investigated. In \citet{Contreras2022b} we found that observed pairs with a mass ratio below 1:10 (i.e. very different masses) are more likely to be close in physical distance than the general population of pairs. For this reason, we believe including these pairs can be important in a study like ours.

In our previous work, we also allowed for groups to be formed, i.e. connecting more than two galaxies. In this case, we will skip that step and create only pairs. This means that if a galaxy A has two close companions B and C, but B and C do not meet the criteria to form a pair between themselves, we will identify this situation as two different pairs: A-B and A-C, rather than a group with the three of them. We use this approach because we are interested now in galaxies being gravitationally bound to each other, not only physically close, and hence it may be counterproductive to connect galaxies like B and C since they do not necessarily have any relation.

Once all the pairs are found, they can be analysed both generally by considering the overall statistics for all the clusters, and also more thoroughly by making use of all the information available in the simulations. We will expand on the latter in the following subsection.

\subsubsection{Classifying 2D pairs as `true' or `false'} \label{sec:method-class}

Using the methodology described above we can associate all the projected galaxies with their `paired' galaxies, taking into account that a galaxy can be paired with more than one galaxy. Then, the 3D information available in simulations can be used to discern if the projected pairs are also close in real space. We focused on this in our first paper \citep{Contreras2022b}, where we differentiated between `good' and `bad' pairs, based on whether the 3D separation between the galaxies was within the allowed 2D range. In the present work, we take a more theoretical approach and, apart from positions, we also use velocity and mass information to determine if the two galaxies are gravitationally bound. 

For this task, we use the prior work done by \citet{Haggar2021}, to which we refer the reader for further information. In this work, galaxy groups are identified in \textsc{The Three Hundred} simulations by determining how many galaxies are associated with each individual galaxy (which is referred to as the `primary' galaxy). Considering each galaxy in the simulation, the other galaxies (referred to as `secondaries') are associated with it if they satisfy certain criteria. First, the total (dark matter, gas and stars) mass of a `secondary' galaxy must be less than that of the `primary' galaxy; and the galaxy must satisfy the condition below:

\begin{equation}
    \frac{v^2}{2}+\Phi(r) < \Phi \left( 2.5 R_{200}^\mathrm{primary} \right)
    \label{eq:roan}
\end{equation}

This condition is the same previously used by \citet{Han2018} and \citet{Choque-Challapa2019} to find galaxy groups. In \Eq{eq:roan}, $\Phi(r)$ represents the gravitational potential due to the primary galaxy at a distance $r$ from its centre, and $v$ is the relative velocity of a secondary galaxy with respect to this primary galaxy. $R_{200}^\mathrm{primary}$ is the radius of the primary galaxy 
(not to be confused with the radius of the main cluster in each of the simulations). If a galaxy is less massive than another one defined as the `primary' and this criterion is satisfied, then the galaxy is considered to be bound to this primary galaxy. 
The advantage of using this kind of definition is that it allows us to include pairs at all points in their orbit -- not only those with distance and velocity separations below some fixed thresholds.

Comparing our identified pairs with the catalogues of gravitationally bound galaxies created by \citet{Haggar2021}, we can check if the galaxies in our pairs are bound or not. In the latter case, the galaxies can be either separate in physical distance, with the pair being a projection effect (as we studied in our previous paper \citealp{Contreras2022b}), or they can be physically close but passing by, not bound to each other (which is called a flyby).
In both cases, the galaxies are expected to evolve independently rather than being merging candidates, and hence we will refer to them as `false' pairs. On the contrary, when an identified pair is also found by \citet{Haggar2021} to be bound, we refer to it as a `true' pair. We will devote the following sections to studying this distinction when applied to all our pairs.

\section{Analysing the results: purity and completeness} \label{sec:stats}

Using the methodology described in the previous section, we can link the galaxies to their close projected companions in each of the 324 clusters. Then, we can classify the identified pairs into `true' or `false' depending on whether the galaxies are gravitationally bound or not. In this section, we will analyse the results obtained and assess how good the observational method to find pairs is when used to detect gravitationally bound galaxies. We will do this first by counting how many of the observed pairs are classified as true (purity), and then by counting how many of the real pairs found in the theoretical work by \citet{Haggar2021} are also found by our methods (completeness).

\subsection{Purity}

\begin{figure}
   \hspace*{-0.1cm}\includegraphics[width=8.5cm]{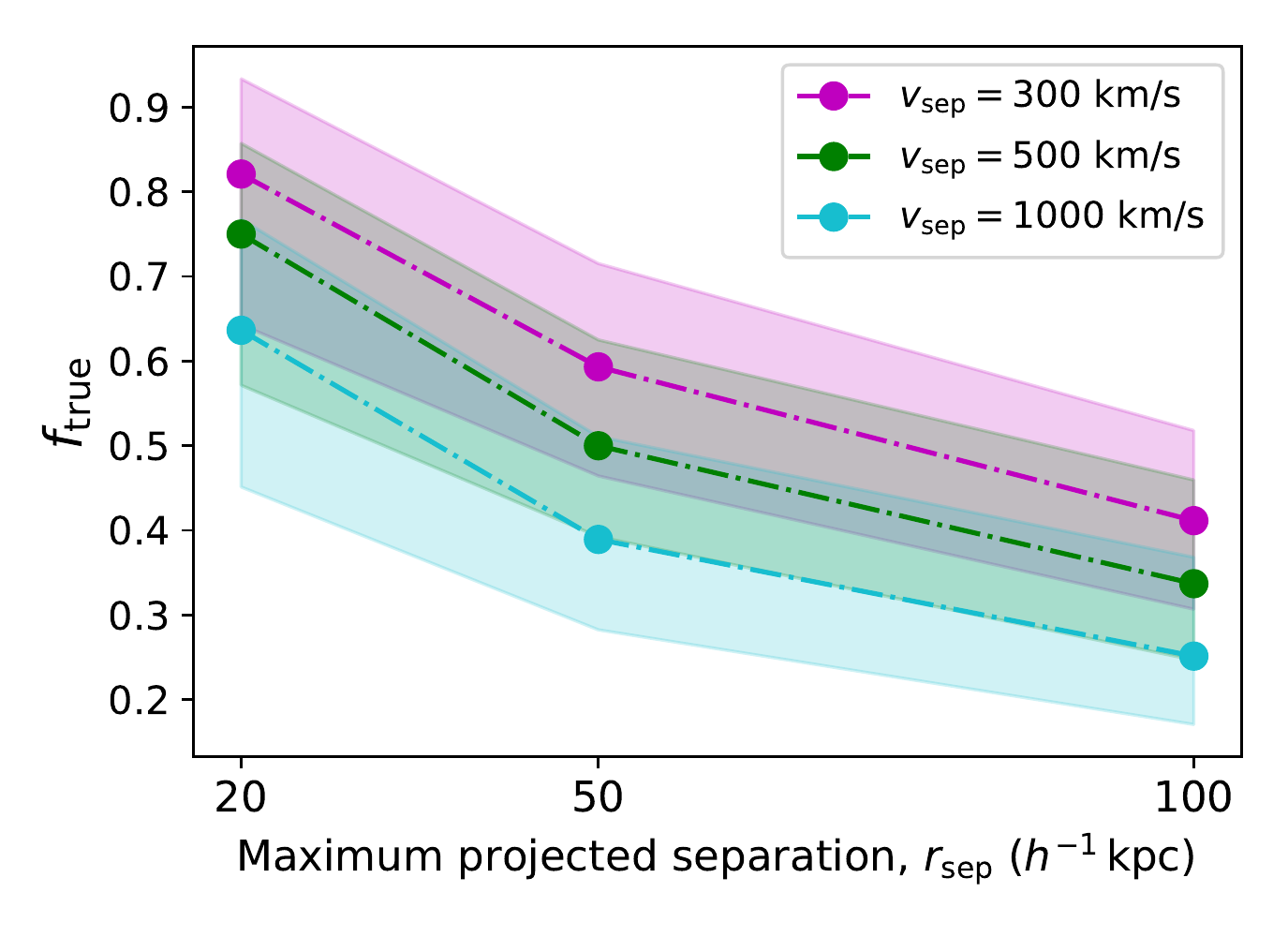}
   \caption{Fraction of the observed pairs that are `true' according to the criteria in \citet{Haggar2021}, i.e purity of the sample. The dots show the median values of the distribution for the 324 clusters in \textsc{The Three Hundred} data set, with the shaded regions indicating the 16th-84th percentiles. The colours indicate the velocity threshold $v_\mathrm{sep}$ used, in magenta 300 km/s, in green 500 km/s and in cyan 1000 km/s.}
\label{fig:purity}
\end{figure}

In this subsection, we study the fraction of the observed pairs that are classified as `true' according to the criteria in \Sec{sec:method-class}. This is the first measure of how good our sample is compared to a true sample, such as the one given by \citet{Haggar2021}. From now on, we will refer to this as the purity of our sample, since it is a measure of the percentage of pairs that are within the required criteria. This value, computed for the nine different combinations of $r_\mathrm{sep}$ and $v_\mathrm{sep}$ is shown in \Fig{fig:purity}. In this figure, the dots show the median values for the 324 clusters (one random projection for each cluster), while the shaded regions show the 16th-84th percentiles. The values in the $x$-axis indicate the different distance separation thresholds, while the colours indicate the different velocity thresholds (as shown in the legend). Through these results, we can see that the purity is very high for the smallest separation of $r_\mathrm{sep}=20$ $\hkpc$, reaching 82 per cent for the most restrictive velocity separation. When increasing the maximum allowed separation, the purity drops significantly. For $v_\mathrm{sep}=300$ km/s, the fraction drops to 59 and 41 per cent for 50 and 100 $\hkpc$ separations respectively, and these values become even lower when relaxing the velocity criterion. In general we can say that, although for a separation of 20 $\hkpc$ the results are very good and the purity is high, increasing this parameter worsens the results significantly. This is especially true when we reach $r_\mathrm{sep}=100$ $\hkpc$, where the purity can be as low as 25 per cent as many pairs are false/not gravitationally bound. Note that purity depends on both spatial separation and velocity separation criteria. 

\subsection{Completeness} 

\begin{figure}
   \hspace*{-0.1cm}\includegraphics[width=8.5cm]{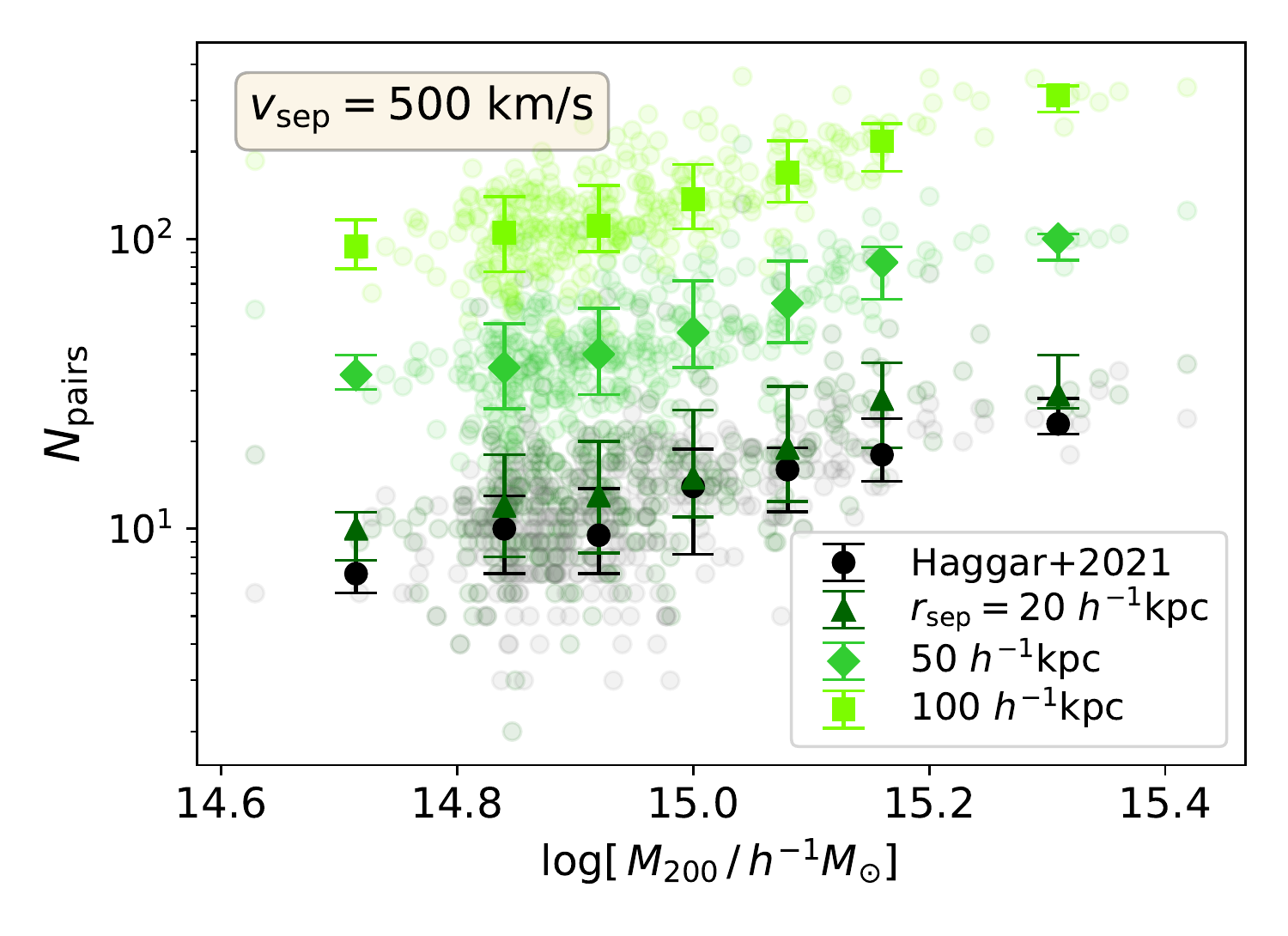}
   \caption{Total number of pairs found for maximum velocity separation $v_\mathrm{sep}=500$ km/s, binning by cluster mass $M_{200}$. From darker to lighter green, the triangles, diamonds and squares show the median values for maximum separations $r_\mathrm{sep}=20$, 50 and 100 $\hkpc$, respectively. The black dots are the values from \citet{Haggar2021}.}
\label{fig:totalpairs}
\end{figure}

\begin{figure}
   \hspace*{-0.1cm}\includegraphics[width=8.5cm]{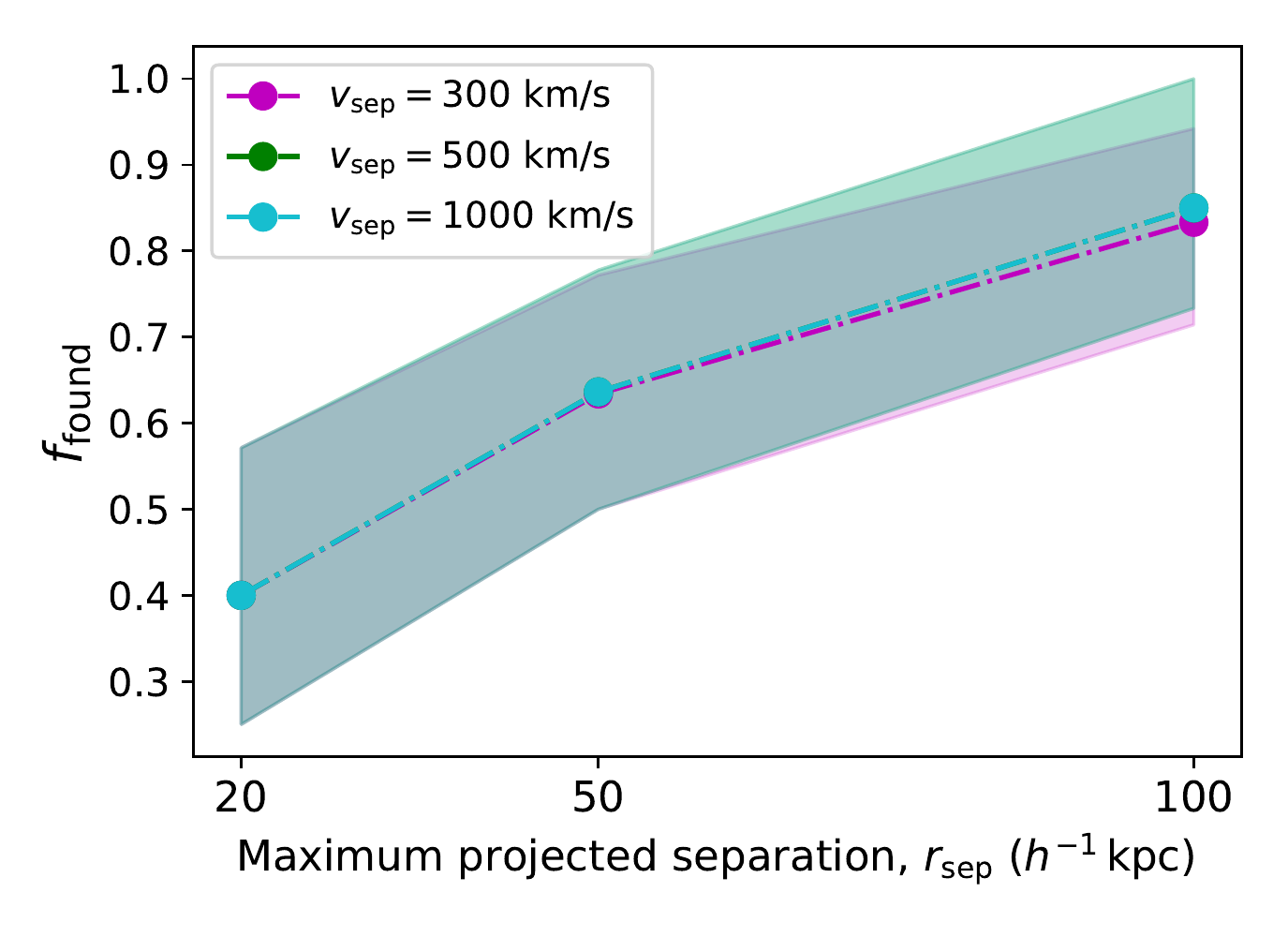}
   \caption{`Pair-completeness' (considering only the pairs from \citealp{Haggar2021}) as a function of maximum separation chosen, $r_\mathrm{sep}$. The dots show the median values of the distribution for the 324 clusters in \textsc{The Three Hundred} data set, with the shaded regions indicating the 16th-84th percentiles. The colours indicate the velocity threshold $v_\mathrm{sep}$ used, in magenta 300 km/s, in green 500 km/s and in cyan 1000 km/s.}
\label{fig:completeness}
\end{figure}

Although purity is a very important parameter to describe the goodness of a sample, it only measures how good the identified pairs are. It does not answer the question of whether all the actual pairs in the simulation are identified with our method or not. We can have very high purity, meaning that all our pairs are true, but at the same time many real pairs that should also have been identified can be missing. The measure of how many of the real pairs are actually found is often referred to as completeness since it measures the degree to which our sample is complete when compared to a `true' sample. 

In our case, it is not easy to compute a precise value for completeness, ranging between 0 and 1, because the true sample we are using from \citet{Haggar2021} contains both pairs and groups of bound galaxies. On the other hand, our sample is designed to contain only pairs, and thus a one-to-one comparison between the two samples is not as straightforward. As a first approach to studying the completeness of our sample, in \Fig{fig:totalpairs} we show the total number of pairs we find as a function of cluster mass $M_{200}$, focusing only on the results for one velocity separation threshold, $v_\mathrm{sep}=500$ km/s. The dots show the median values for all the clusters in each mass bin and the error bars indicate the 16th-84th percentiles. From darker to lighter green, the dots show the results for 20 (triangles), 50 (diamonds) and 100 $\hkpc$ (squares) separations respectively. Black dots are the values for \citet{Haggar2021}, noting that we count only the pairs and not the groups, so that we make a comparison just between pairs. Apart from the expected trend for more massive galaxy clusters to have more pairs (both theoretically and observationally), we see in this plot that we are finding many more pairs than in the `true' sample. This is especially the case for the higher separation thresholds $r_\mathrm{sep}=$ 50 and 100 $\hkpc$, which can explain the low purity we saw in \Fig{fig:purity} for these separations. Even for $r_\mathrm{sep}=20$ $\hkpc$ we see that our values are slightly higher than the black dots. One thing to keep in mind is that we are removing the groups from \citet{Haggar2021} for this plot, while in our sample we are allowing each galaxy to be in more than one pair. Including all the groups would raise the black dots a little, but the values from this work would remain significantly higher.

Although we have seen in \Fig{fig:totalpairs} that we are finding too many pairs, the question still remains of whether we are finding all the real pairs or not. To address this issue, we define a pseudo completeness using only the pairs from \citet{Haggar2021}, which we will call `pair-completeness'. This can be computed as the fraction of pairs from the true sample that are found by our methods. Although it is not the full completeness since we are ignoring part of the real sample, it is still a measure of how complete our different samples are -- how good they are at finding all the real pairs.

Similarly to \Fig{fig:purity}, \Fig{fig:completeness} shows the pair-completeness for the nine different combinations of $r_\mathrm{sep}$ and $v_\mathrm{sep}$, the dots being the median values and shaded regions the 16th-84th percentiles. The first thing to note here is that there is no dependence on the $v_\mathrm{sep}$ threshold, with the values being almost the same for the three different thresholds selected. This means that bound galaxy pairs always have a line-of-sight velocity separation below $\sim 300$ km/s, so increasing this threshold simply results in the inclusion of more false pairs in the sample (which can be either far away galaxies or flybys). For future studies, this indicates there is no reason to use a $v_\mathrm{sep}$ higher than 300 km/s, and lowering this limit should even be considered in order to increase purity without a significant loss in completeness. 

Regarding the values of the pair-completeness themselves, we see in \Fig{fig:completeness} that for $r_\mathrm{sep}=20$ $\hkpc$ the completeness is only 40 per cent now. In other words, although the purity is very high with this restricted separation threshold (reaching 85 per cent), we miss a very relevant amount of true pairs. For the higher values of $r_\mathrm{sep}$, 50 and 100 $\hkpc$, pair-completeness increases significantly, reaching 63 and 83 per cent respectively. Interestingly, completeness seems only to depend on spatial separation with very minor changes with different velocity separations. This means that the velocity separation thresholds used here are still too large to enter the gravitationally bound calculation.

To summarise the results in this section, we have shown that a compromise needs to be found between the purity and completeness of a sample. If the selection criteria are very restrictive (as for $r_\mathrm{sep}=20$ $\hkpc$), the purity will be high but the sample might not be complete enough. On the contrary, more relaxed criteria like $r_\mathrm{sep}=100$ $\hkpc$ can lead to a less pure sample that contains a much more significant fraction of the real pairs. Keeping this in mind, the definition of close pair can be chosen according to the particulars and aims of each specific study.

\section{Improving the classification} \label{sec:results}

In the previous section, we showed how the different thresholds adopted when finding galaxy pairs affect the quality of the obtained sample. This way, for the most restrictive values of $r_\mathrm{sep}$ and $v_\mathrm{sep}$ we obtained a high purity (majority of true pairs) at the cost of low completeness (many real pairs missing). On the other hand, for less restrictive criteria, we found a low purity but with high completeness. This is already valuable information which can help to choose the desired parameters depending on the particulars of the study to be done. However, in general, it would be desirable to have a method that maximises both purity and completeness, so that observational samples of pairs can be created to be used for multiple applications and show a good agreement with theory. In this section, we present our approach to this issue, by applying a machine learning model to classify observed galaxy pairs based on their known properties.

\subsection{Random Forest Algorithm} \label{sec:rf_intro}
Using machine learning has the advantage that it allows us to work with very big data sets, and analyse amounts of data that might be very difficult -- if not impossible -- to inspect manually. In our case, this means we can use all the available properties of the galaxies rather than having to select the ones we believe to be more important. We hence reduce possible biases in the results related to this. In simulations, where a large amount of information is available, this can be of special relevance.

For our specific problem, namely classification into two classes (true or false), one algorithm that is widely used due to its simplicity, stability and robustness is the random forest. Random Forest (RF; \citealp{Breiman2001}) is a commonly used machine learning algorithm constructed by combining multiple decision trees. A decision tree is a tree-like graph constructed top-down from a root node. Each node partitions the data into two subsets based on the values of the input parameters. The resulting leaf nodes can either be a new node or a final prediction. At each step, the best split is chosen based on minimising the Gini impurity -- this is a measure of the likelihood of a random data point being misclassified if it were given a random class label based on the class distribution in the data set. A Gini impurity of 0 can only be achieved if the split perfectly separates the data points into the two given classes.

After splitting the data into a training and a test data set, different random subsamples of the training set are used to construct a number of decision trees. When using it for classification, the output of the RF is the class selected by most individual trees. If applied to the test set, the output can be compared to the ground truth, thus evaluating the performance of the model. This algorithm can also output the importance of each feature used for the classification. This way, as well as directly using the model to classify pairs, we can learn which features are the most important when discerning if an observed pair is gravitationally bound or not.

In the following subsections, we explain in detail how we apply this algorithm to different subsets of our data, with the aim to find a model that classifies observed pairs into true and false with improved performance. We start with a theoretical approach including many different properties of the involved galaxies, and then use the results to make a selection of those properties that appear to be the most important and can be used in observational studies to classify pairs. 
A more in-depth description of the models, together with the validation of their performance, is shown in Appendix \ref{appendix:ml_validation}.

\subsection{Theoretical approach: all properties} \label{sec:rf_theoretical}

As a first approach to our problem, we start from a more theoretical point of view, including information that is only available in simulations. This will help us to understand the performance of our method and the data itself. The main goal of this subsection is to find which properties of the galaxies are most important when determining whether they belong to a bound pair. We start by describing the data used as inputs for the random forest algorithm to perform the classification task, i.e. a list of properties for all the galaxies included in the pair sample. 

Given the already mentioned advantage of machine learning in handling large amounts of data, we blindly use all the properties given by AHF, combined with those given by \textsc{Caesar}. For AHF, these include a set of properties regarding all the particles in each halo, i.e. dark matter, stars and gas. The full list (60 properties from AHF) can be seen in \Tab{table:allprops_ahf} in Appendix \ref{appendix:props}, but in general, they are properties related to mass, radius, velocity, spin parameter, angular momentum, the moment of inertia tensor and kinetic and potential energies. Then, AHF can also repeat these calculations but using only one family of particles. In this case, we include also the same properties but only for the star particles, including also the mean metallicity and the stellar-to-halo mass ratio. By applying the \textsc{stardust} code (see \Sec{sec:data} for further detail), we also compute luminosities and magnitudes in different Johnson bands and use them to compute different colour indices.

Regarding the \textsc{Caesar} properties, they are listed and briefly described in \Tab{table:allprops_caesar} (55 properties from \textsc{Caesar}). They can also be separated into different groups: masses computed for different types of particles, i.e. gas, stars and both together, and for different apertures; as well as radii, angular momentum and velocity dispersion for the different components. Then, \textsc{Caesar} also includes the option to compute luminosities in different spectral bands, and hence colours. Finally, we also include age, metallicity and star formation rate (SFR) for all the galaxies. Although there is some overlap between AHF and \textsc{Caesar} properties, we decide to keep all of them, since this method allows us to include as much information as possible without any extra cost. While AHF properties are theoretically oriented, \textsc{Caesar} properties are more observationally oriented. Additionally, the same properties are obtained in different ways in the two catalogues, and so this way we will also be able to check for consistency between them. 

We collect all these properties for all the galaxies that appear in our sample of pairs. Then, there are two ways in which we can use them for our purpose. We can either study the properties of the galaxies individually, or the properties of the pairs of galaxies, i.e. the properties of one galaxy but in relation to the other galaxy in the pair.

\subsubsection{Properties of galaxies in pairs} \label{sec:rf_props}

For this section, we select the pairs found using the thresholds in projected distance $r_\mathrm{sep}=100$ $\hkpc$ and in velocity separation $v_\mathrm{sep}=500$ km/s. For the velocity, we saw in \Fig{fig:completeness} that the pair-completeness is independent of the threshold used, so we simply select the middle value of 500 km/s. Regarding the distance separation, we start with 100 $\hkpc$ because it can be seen in \Fig{fig:purity} that this separation is where the most improvement can be made in terms of purity of the classification.

We first work with the properties of the individual galaxies in the identified pairs. This means that we use the different features of both galaxies in every pair but as two independent galaxies, rather than considering them two attributes of the same pair. This way we include all the properties mentioned before (together with whether the galaxy belongs to a true or false pair) as inputs for the random forest, which we apply using the function \texttt{RandomForestClassifier} from the \textit{scikit-learn} library\footnote{\url{https://scikit-learn.org/}} \citep{scikit-learn}. We use 70 per cent of the whole data set for training and the remaining 30 per cent for testing. 
After training this model with our data, we can obtain the relative importance of each feature when classifying the pairs into `true' or `false'. This is computed for each feature as an average over all the decision trees, based on how splitting the tree about this feature affects the purity of the results. 

\begin{figure*}
  \hspace*{-0.0cm}\includegraphics[width=13.4cm]{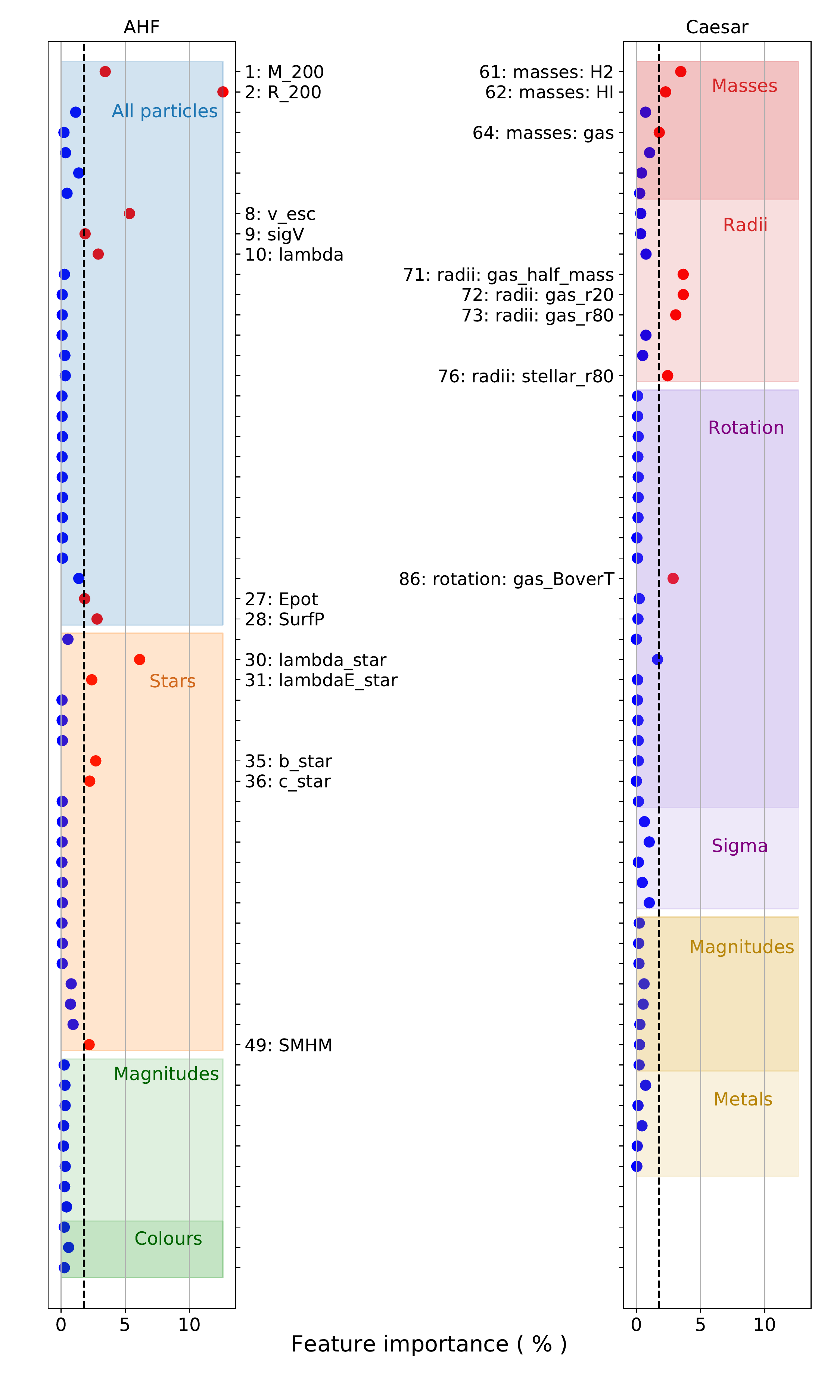}
  \vspace*{-0.4cm}
  \caption{Feature importance in percentage as given by the random forest algorithm in \Sec{sec:rf_props}, for all the AHF (left) and \textsc{Caesar} (right) \textbf{properties of the individual galaxies} combined. Red dots designate the 20 most important features. For easier visualization, the labels are only shown for these properties, but a list with all of the properties together with a brief description of them can be seen in Tables~\ref{table:allprops_ahf} and~\ref{table:allprops_caesar}. The different coloured regions in the plot separate related or similar properties, again only for visualization purposes. The vertical dashed line corresponds to the importance of the 20th most important feature.}
\label{fig:rf_props_ahf_caesar}
\end{figure*}

The importance of each property is shown in \Fig{fig:rf_props_ahf_caesar}. The values are in percentages so that the sum of all properties is equal to 100. The red dots correspond to the 20 most important properties, whose labels can be seen beside, while the blue dots correspond to the rest of the properties, whose names are not shown for clarity (but can be seen in Tables~\ref{table:allprops_ahf} and~\ref{table:allprops_caesar}). For easier visualization, the plot is separated into two columns, AHF properties in the left and \textsc{Caesar} properties in the right, but we note that the algorithm has been applied to both of them simultaneously. 
In general, we see that the most important feature is $R_{200}$, followed by escape velocity, mass and spin parameter. Since the mass is defined as mass enclosed within $R_{200}$, and the escape velocity is computed based on both the mass and the radius, these three quantities in the end contain very similar information. 
The spin parameter however depends also on the angular momentum of the object, so it provides different information, indicating that the spin is an important parameter to distinguish whether a galaxy belongs to a true pair.
Another relevant feature is the surface pressure of the galaxies, which is computed in AHF following the \citet{Shaw2006} definition, and accounts for the particles that are bound to the halo but at its boundary, so that it is also related to the size and mass of the objects. For \textsc{Caesar} we see that radii are important too, mainly the gas radius in this case, but also for the stars. We also want to highlight the importance of the gas mass given by \textsc{Caesar}, which we find to be an interesting result (and will further investigate in the following subsection).

Regarding the AHF properties including only the stellar particles, we see again that the spin parameter shows up, and in this case also the $b$ and $c$ parameters. These are defined as the second (for $b$) or third (for $c$) largest axis of the moment of inertia tensor divided by the largest one. They are a measure of the shape of the galaxy, with a value equal to 1 indicating perfect sphericity. The stellar-to-halo mass ratio is also shown to be relevant.
We see that the magnitudes in all the different bands do not play an important role in this classification. The AHF colours are shown to be slightly more relevant than the magnitudes themselves but still with low feature importance compared to other properties.

In general we can say that, when talking about properties of individual galaxies in observed pairs, the size of the galaxies together with their mass, spin parameter, gas content, stellar-to-halo mass ratio and shape of their stellar component are the properties that help the most to distinguish between true and false pairs. 
We also note here that, as its name implies, RF has a random component, so that the specific values shown in \Fig{fig:rf_props_ahf_caesar} will only be repeated to a certain accuracy in different realisations. Thus, we prefer to emphasize the order of importance of the different features, rather than the values of importance themselves.
Moreover, the resulting feature importances of the RF can be affected if there are some features that are strongly correlated, like $M_{200}$ and $R_{200}$. This is because in general the model has no preference for one over the other, and thus their values of importance can be reduced. However, this effect does not affect the order of importance of the features, or the distinction between important and non-relevant variables \citep{Genuer2010}, and thus it is not an issue for our results here.

At this point we also studied the possibility of one of the galaxies in the pair being more important than the other one, to see if, for instance, the most massive (or primary) galaxy, is the one leading the classification criteria. For this, we separated the galaxies of each pair into the most and the least massive of them ($\mathrm{G}_1$ and $\mathrm{G}_2$ respectively) and considered their properties as two different properties of the same pair. We then applied the RF algorithm and studied the resulting feature importance. In general (the plots are not shown here for clarity) we saw that $\mathrm{G}_1$, i.e. the most massive galaxy, dominates, with a summed importance of its features of $\sim 57$ per cent against 43 for $\mathrm{G}_2$. When looking at the different features individually, we see that $\mathrm{G}_1$ is more important in most of the features except for the radius, which, as can be seen in \Fig{fig:rf_props_ahf_caesar}, is actually the most important property. Therefore, $R_{200}$ of the less massive galaxy is significantly more important than that of the most massive galaxy when making this classification. The rest of the properties, like gas content and $b_*$ and $c_*$ (which are the same as $b$ and $c$ but computed only for the stellar particles), show a predominance of $\mathrm{G}_1$, while for the spin parameter both galaxies have a similar relevance.

\subsubsection{Properties of galaxy pairs} \label{sec:rf_ratios}
After studying the properties of individual galaxies in the pairs, we move to analysing properties of the pairs themselves, by computing the ratio of the properties between the two pair members. That is, for each property $P$ used previously, we compute the ratio $P_2/P_1$, where $P_2$ and $P_1$ are the specific properties of the two galaxies in the pair, chosen so that this ratio is always $\leq 1$.

\begin{figure*}
  \hspace*{-0.0cm}\includegraphics[width=13.2cm]{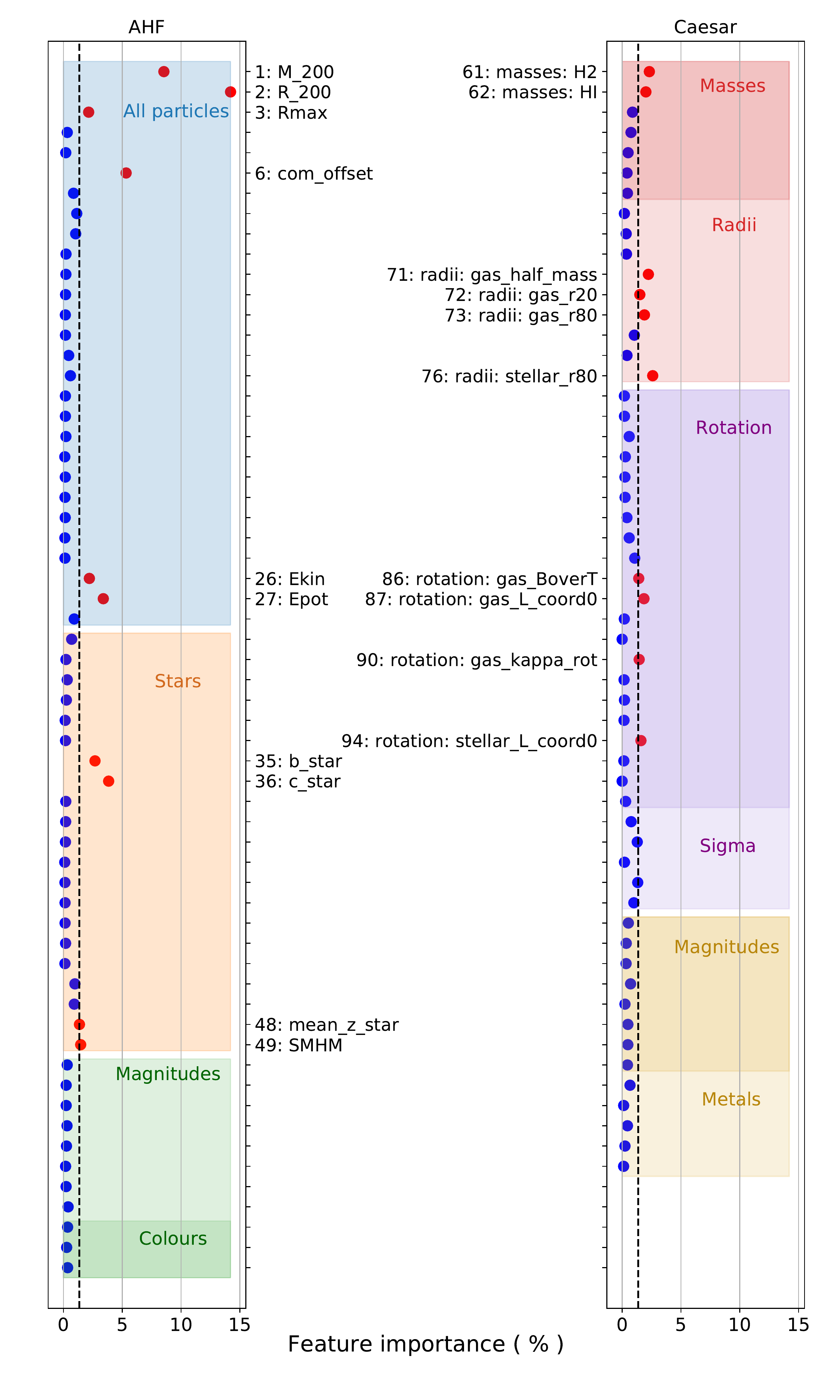}
  \vspace*{-0.4cm}
  \caption{Feature importance in percentage as given by the random forest algorithm in \Sec{sec:rf_ratios}, for the AHF (left) and \textsc{Caesar} (right) properties of the pairs of galaxies, obtained as the \textbf{ratio of each property between the two pair members}, $P_2/P_1$, so that $P_2/P_1 \leq 1$. Red dots designate the 20 most important features. For easier visualization, the labels are only shown for these properties, but a list with all of the properties together with a brief description of them can be seen in Tables~\ref{table:allprops_ahf} and~\ref{table:allprops_caesar}. The different coloured regions in the plot separate related or similar properties, again only for visualization purposes. The vertical dashed line corresponds to the importance of the 20th most important feature.}
\label{fig:rf_ratios_ahf_caesar}
\end{figure*}

Similarly to \Fig{fig:rf_props_ahf_caesar}, in \Fig{fig:rf_ratios_ahf_caesar} we show the feature importance when applying the RF algorithm to classify the identified pairs instead. In this case, comparing with \Fig{fig:rf_props_ahf_caesar}, we see that a similar selection of features appear to be relevant: virial radius, mass and the centre of mass offset. The shape of the stellar component remains relevant, indicating that the relation between the shapes of the two galaxies is also important, not just the shapes of the galaxies alone. The same is true of the stellar-to-halo mass ratio. These results can indicate a possible stripping of the stars and dark matter halo of one galaxy by the other, a situation we already described in our previous paper \citet{Contreras2022b}. There we concluded that galaxies in physically close pairs showed a tendency to have different shapes and stellar-to-halo mass ratios, as opposed to galaxies in spurious pairs created by projection effects, where the galaxies could be more similar in both parameters. The relevance of the mean stellar metallicity ratio here also confirms the results in \citet{Contreras2022b}. 

The main difference with \Fig{fig:rf_props_ahf_caesar} is in the spin parameter. While previously we saw that it was quite important, both for all particles and only for stars, we see now that the ratio of the two galaxies in a pair does not seem to be relevant. This can be interpreted to mean that, although galaxies in true pairs tend to have specific values of the spin parameter, the relation between the two of them is not affected by the two galaxies forming a gravitationally bound pair.

Finally, regarding the gas content, we see again that it remains relevant, for the ratio of both $\mathrm{H_2}$ and HI components. Although here we only list the most relevant features as highlighted by the RF algorithm, we will elaborate on them and how they can be used to classify pairs in the following subsections. Hence we emphasise that the purpose of these models with numerous properties, rather than the classification itself, is to find the most relevant features when classifying galaxy pairs, so that we can now study and work with a more reasonable set of properties.

We want to highlight again that the results shown in these two subsections have been obtained for the pairs found using the thresholds $r_\mathrm{sep}=100$ $\hkpc$ and $v_\mathrm{sep}=500$ km/s. Although we only show these results here, we have also repeated these calculations for $r_\mathrm{sep}=50$ $\hkpc$. We have seen that, in spite of some differences due to the random component of this analysis, the results obtained are essentially the same regardless of the thresholds selected.

\subsection{Observational approach: selection of properties} \label{sec:rf_observational}

In the previous subsection we blindly included all the available AHF and \textsc{Caesar} properties as input for the RF algorithm to predict if an observed pair was gravitationally bound or not. We used this model to determine which properties of the pairs and the involved galaxies played a more important role when doing this classification. Although we obtained relevant information, this was a very theoretical approach, since real observational studies do not provide that much information about the galaxies. For this reason, in order to have something useful from an observer's perspective, we will now make a selection of properties that are more easily accessible to observers. This way, we will train a new RF only on a selection of relevant properties of the galaxies. We will also try to understand the physical situation behind the ML model, by investigating how the selected properties differentiate the true from the false pairs.

Based on the results described in \ref{sec:rf_theoretical} and the plots of feature importance in Figs~\ref{fig:rf_props_ahf_caesar} and~\ref{fig:rf_ratios_ahf_caesar}, we can make a selection of properties that can be available in observations and at the same time are relevant for the interests of this work. We also aim to reduce the number of properties used as much as possible, thus making this method more realistic and manageable. Consequently, in this subsection we decide to work with the following properties of each pair:
\begin{itemize}
    \item Stellar spin parameter, $\lambda_*$. We decide to keep this value for both galaxies in the pair, since we saw in \Sec{sec:rf_props} that they were similarly important. The ratio of the two values was not found to be as important in \Sec{sec:rf_ratios}, so we do not select it.
    \item Stellar shape parameter, $c_*$, defined as the ratio of the minor to major axis of the moment of inertia tensor. Based also on our previous results, we keep only that of the most massive galaxy (as we find the ratio $c_\mathrm{*,1}/c_\mathrm{*,2}$ to be less important than $c_\mathrm{*1}$).
    \item Molecular gas content $M_\mathrm{H_2}$, only of the most massive galaxy in the pair.
    \item Radius enclosing 80 per cent of the stars, $R_\mathrm{*,80}$. In this case we keep the values for both galaxies, which are shown to both be important.
    \item Mean stellar metallicity ratio, $Z_\mathrm{*,2}/Z_\mathrm{*,1}$. 
\end{itemize}

We thus select a list of seven different properties characterising each galaxy pair: $\lambda_\mathrm{*,1}$, $\lambda_\mathrm{*,2}$, $c_\mathrm{*,1}$, $M_\mathrm{H_2,1}$, $R_\mathrm{*,80,1}$, $R_\mathrm{*,80,2}$ and $Z_\mathrm{*,2}/Z_\mathrm{*,1}$.

\subsubsection{Application of Random Forest} \label{sec:rf_obs_1}

Given this selection of properties, we can train a new random forest model that uses only this selected information to classify the pairs into true or false (gravitationally bound or not). Comparing the output classification to the `ground truth' in \citet{Haggar2021} (see \Sec{sec:method-class}), we can compute the purity and completeness of this classification similarly to how we did in \Sec{sec:stats}. In this case, using the seven properties previously listed, we obtain a purity of 82 per cent (i.e. from all the pairs classified as `true', 82 per cent of them were actually true), as opposed to the previous 34 per cent in \Sec{sec:stats} (see \Fig{fig:purity} for $r_\mathrm{sep}=100$ $\hkpc$ and $v_\mathrm{sep}=500$ km/s), demonstrating a very significant improvement. Regarding completeness, the previous value we had of 85 per cent (see \Fig{fig:completeness}) is now reduced to 50 per cent. This means the random forest classifier correctly classifies as `true' 50 per cent of the `observed' true pairs. Although this is a non-negligible decrease, it is still a high value for completeness, especially if we consider the remarkable increase in purity. For instance, for $r_\mathrm{sep}=20$ $\hkpc$ and $v_\mathrm{sep}=300$ km/s we had a very similar purity (82 per cent) but with a lower completeness (40 per cent, see \Sec{sec:stats}).

Apart from the seven selected properties together, we also consider different combinations of them, in order to create more realistic models that can be applied when only some of these features are available. For instance, we find $c_\mathrm{*,1}$ to be the most important of them, and hence we try a model that uses only this attribute. Similarly, we train four more distinct models with combinations of the selected properties. We summarize all the models here: 
\begin{enumerate}
    \item[\textbf{(A)}] All seven selected properties
    \item[\textbf{(B)}] Only stellar shape $c_\mathrm{*,1}$
    \item[\textbf{(C)}] Only gas content $M_\mathrm{H_2,1}$
    \item[\textbf{(D)}] Stellar shape and radii: $c_\mathrm{*,1}$, $R_\mathrm{*,80,1}$ and $R_\mathrm{*,80,2}$
    \item[\textbf{(E)}] Gas content and radii: $M_\mathrm{H_2,1}$, $R_\mathrm{*,80,1}$ and $R_\mathrm{*,80,2}$
    \item[\textbf{(F)}] Stellar spin parameter, gas content and radii: $\lambda_\mathrm{*,1}$, $\lambda_\mathrm{*,2}$, $M_\mathrm{H_2,1}$, $R_\mathrm{*,80,1}$ and $R_\mathrm{*,80,2}$
\end{enumerate}

The performance of the different models can be seen in the first row of \Tab{table:rf_results}. For $r_\mathrm{sep}=100$ $\hkpc$ and $v_\mathrm{sep}=500$ km/s, the first two columns show the values of purity and pair-completeness as described in \Sec{sec:stats}, with the difference that, in Figs~\ref{fig:purity} and~\ref{fig:completeness} we computed the value for each cluster and then showed the median values, while in \Tab{table:rf_results} we are directly showing the values computed by stacking together the pairs from all the clusters. The following columns show the performance of the different RF models when trained and tested with the different samples (we use 70 per cent of the data for training and the remaining 30 per cent for testing). We see that the model using only $c_\mathrm{*,1}$ (model B) already gives 72 per cent purity and 32 per cent completeness, while combining this with the shape and radii information (model D) raises the purity to 77 per cent and completeness to 42 per cent. Additionally, the gas content alone (model C) yields a high purity (71 per cent) but with quite low completeness (29 per cent), but these values can be improved if we include the radii information (see model E). In model F, apart from gas content and radii, we also include the spin parameter of the galaxies, raising the purity and completeness to values considerably closer to the model with all properties. In general, any combination of three different properties shows a very similar performance to model F. It is important to keep in mind that, although columns (A)-(F) in \Tab{table:rf_results} refer to the performance of the RF algorithm itself, this is trained on the `observational samples' (obtained following the methodology in \Sec{sec:method}), so the completeness here does not use the total number of pairs from \citet{Haggar2021}, but rather the true pairs found with the observational method. Given that the C for each sample is the maximum value that the ML method can achieve, we introduce relative completeness, $c$, which is a further renormalization of the C value of each sample.

As before, we can repeat the results in the first row of \Tab{table:rf_results}, but changing the paired sample used to that obtained using the different $r_\mathrm{sep}$ - $v_\mathrm{sep}$ combinations. The performance of the different models can be seen in the following rows of \Tab{table:rf_results}. For $r_\mathrm{sep}=20$ $\hkpc$, the purity is already high with the observational model, but we see that the different random forests, even those that require only one or two properties as input, achieve purity around 90 per cent with still very high completeness. For $r_\mathrm{sep}=50$ $\hkpc$ the behaviour of the values is similar to that in the first row, with a performance that clearly improves on the observational one, reaching more than 80 per cent purity with more than 70 per cent in completeness.

\begin{table*}
\centering
\caption{For each combination of projected distance and line-of-sight velocity separation thresholds ($r_\mathrm{sep}$ and $v_\mathrm{sep}$), fraction of `true' pairs (purity, P) and fraction of pairs from \citet{Haggar2021} that are found using the observational method to identify pairs (completeness, C). The following columns indicate the resulting purity and completeness when training and testing a random forest (RF) algorithm with the selected properties of the galaxies and pairs in the given observational samples. Note that C (observational) is the absolute completeness, while c for the different models is the relative completeness of the model itself, renormalised taking into account that C is the maximum value that can be achieved. The properties used in each of the models are listed in the main text in \Sec{sec:rf_obs_1}.}
\begin{tabular}{r|c|c|c|c|c|c|c|c|c}
\hline
 {$\bm{r_\mathrm{sep}}$} &     $\bm{v_\mathrm{sep}}$        & \multicolumn{2}{c}{\textbf{Observational}} & {\textbf{RF: (A)}} & {\textbf{(B)}} & {\textbf{(C)}} & {\textbf{(D)}} & {\textbf{(E)}} & {\textbf{(F)}} \\
{{($\hkpc$)}} & {(km/s)}  & P & C & p -- c & p -- c & p -- c & p -- c & p -- c & p -- c \\ \hline \hline
\textbf{100}  & \textbf{500}    & 0.354  & 0.836     & 0.817 -- 0.502     & 0.717 -- 0.323     & 0.714 -- 0.293     & 0.770 -- 0.420     & 0.751 -- 0.402     & 0.783 -- 0.504           \\ \hline
\textbf{20}  & \textbf{300}    & 0.718  & 0.403      & 0.929 -- 0.897     & 0.815 -- 0.886     & 0.788 -- 0.913     & 0.918 -- 0.878     & 0.917 -- 0.883     & 0.924 -- 0.888            \\ 
             & \textbf{500}    & 0.656  & 0.406      & 0.900 -- 0.876     & 0.771 -- 0.823     & 0.699 -- 0.892     & 0.885 -- 0.853     & 0.878 -- 0.846     & 0.898 -- 0.868            \\ 
             & \textbf{1000}   & 0.564  & 0.406      & 0.908 -- 0.826     & 0.748 -- 0.650     & 0.854 -- 0.419     & 0.872 -- 0.807     & 0.865 -- 0.796     & 0.898 -- 0.819            \\ \hline 
\textbf{50}  & \textbf{300}    & 0.578  & 0.627      & 0.866 -- 0.809     & 0.751 -- 0.653     & 0.824 -- 0.478     & 0.817 -- 0.785     & 0.822 -- 0.771     & 0.846 -- 0.818            \\ 
             & \textbf{500}    & 0.499  & 0.634      & 0.873 -- 0.727     & 0.739 -- 0.534     & 0.785 -- 0.432     & 0.830 -- 0.682     & 0.828 -- 0.661     & 0.853 -- 0.731            \\
             & \textbf{1000}   & 0.393  & 0.634      & 0.830 -- 0.619     & 0.708 -- 0.408     & 0.760 -- 0.370     & 0.786 -- 0.565     & 0.778 -- 0.556     & 0.808 -- 0.616            \\ \hline 
\textbf{100} & \textbf{300}    & 0.419  & 0.825      & 0.821 -- 0.609     & 0.734 -- 0.421     & 0.737 -- 0.378     & 0.776 -- 0.548     & 0.762 -- 0.542     & 0.805 -- 0.605            \\ 
             & \textbf{1000}   & 0.274  & 0.836      & 0.816 -- 0.351     & 0.702 -- 0.177     & 0.697 -- 0.186     & 0.732 -- 0.291     & 0.735 -- 0.269     & 0.797 -- 0.345      \\ \hline
\end{tabular}
\label{table:rf_results}
\end{table*}

\subsubsection{Interpretation of the results} \label{sec:rf_obs_2}

In the previous subsection we have applied a ML algorithm to classify galaxy pairs into gravitationally bound or not. However, this was done as a `black box', in the sense that there was no physical interpretation of the results, we did not address the question of why the selected properties were important when making this classification. In this subsection we will try to give some insight into this and understand the physical situation behind these classification models. For that, we individually analyse each of the properties previously selected: spin and shape of the stellar component, gas content, radius containing 80 per cent of the stars and mean stellar metallicity. 

\begin{figure}
   \includegraphics[width=8.6cm]{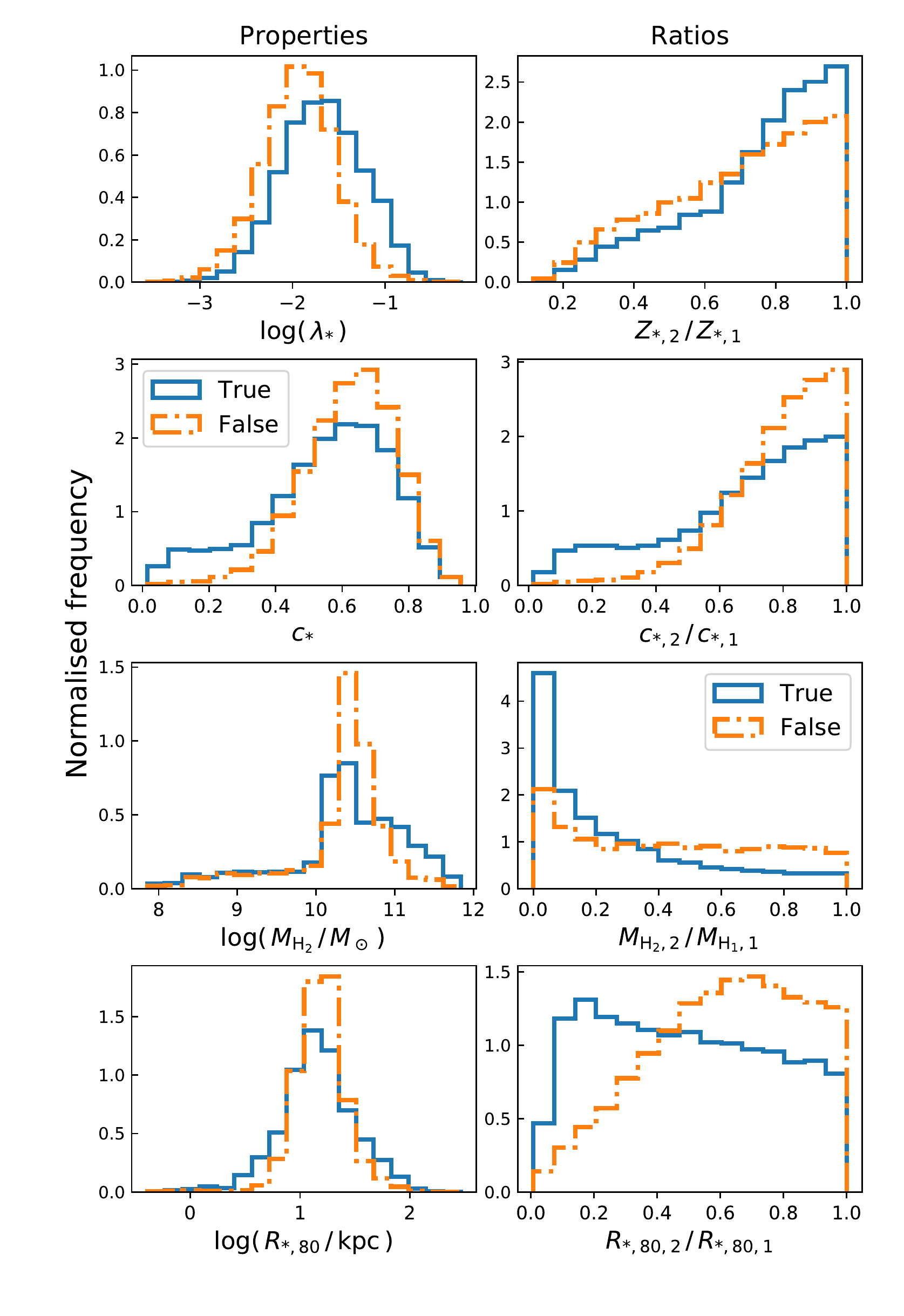}
   \vspace{-0.6cm}\caption{\textbf{Left column,} distribution of the value of the given property (from top to bottom: stellar spin parameter, stellar shape, $\mathrm{H_2}$ mass and radius $R_\mathrm{*,80}$) for all the galaxies in the paired sample, separated into those belonging to `true' (solid blue lines) and `false' (dash-dotted orange) pairs. \textbf{Right column,} distribution of the ratio of the given property (mean stellar metallicity, stellar shape, $\mathrm{H_2}$ mass and radius $R_\mathrm{*,80}$) between the two galaxies in the pair.}
\label{fig:rf_obs_2}
\end{figure}

In the left column of \Fig{fig:rf_obs_2} we show the distribution of the different properties of the individual galaxies, separating them into those belonging to true (solid blue lines) and false pairs (dash-dotted orange lines). As before, these results have been obtained for the pair sample with $r_\mathrm{sep}=100$ $\hkpc$ and $v_\mathrm{sep}=500$ km/s, but the same conclusions hold for different thresholds. From top to bottom, the properties shown in \Fig{fig:rf_obs_2} are stellar spin parameter, $\lambda_\mathrm{*}$, shape of the stellar component, $c_{*}$, $\mathrm{H_2}$ gas mass $M_\mathrm{H_2}$ and radius containing 80 per cent of the stars, $R_\mathrm{*,80}$. For these plots we do not make a distinction between the two galaxies in the same pair, they are both included in the same distribution. Comparing the distributions for true and false pairs, although the differences between them are small, we see that galaxies in true pairs tend to have a higher spin parameter and gas content, while the shape parameter is lower, indicating that these galaxies are less spherical. Regarding the radius, we see that the distribution for true pairs is wider, reaching both higher and lower values than that of the false pairs.

The right column of \Fig{fig:rf_obs_2} shows the ratio of the given properties between the two galaxies in a pair. These properties are the same as in the left column, except for the spin parameter, which we saw in \Sec{sec:rf_ratios} was not relevant for this separation, and instead we include the mean stellar metallicity. We can see that for the metallicity, true pairs are more likely to have galaxies with similar values of metallicity, i.e. $Z_\mathrm{*,2}/Z_\mathrm{*,1}$ close to 1, than false pairs. For the shapes, the situation is the opposite, pairs with $c_\mathrm{*,2}/c_\mathrm{*,1} \sim 1$ are more likely to be false, and those where the galaxies have very different shapes are almost all true pairs. The conclusion is similar for the gas content and the radius: physically bound pairs have galaxies with different properties, for instance one with high $R_\mathrm{*,80}$ and one with a lower value. 

The results in \Fig{fig:rf_obs_2} can be interpreted as a sign of the interaction between the two galaxies in the pair: the spin parameter is growing because the galaxies are physically bound and getting closer together, which is also affecting their shape. One of the galaxies can become very elongated due to this, while the other galaxy remains with a more spherical shape, explaining the situation we see for the ratio. 

For the gas content, we have to be careful because actually most of the galaxies have no gas content. In \Fig{fig:rf_obs_2} we are only including the pairs where both of the galaxies have gas. In fact, around 80 per cent of the galaxies in the paired sample have no $\mathrm{H_2}$. In turn, 70 per cent of these galaxies are in false pairs and only 30 per cent of them are in true pairs. This means that if a galaxy without $\mathrm{H_2}$ is in a pair, this is most likely a false pair (70 per cent likelihood against the general 65 per cent of false pairs for this sample). In \Fig{fig:rf_obs_2} (left column, third row) we see that galaxies in true pairs tend to have more gas than those in false pairs. This is simply the continuation of the previous trend: the higher the gas content in the galaxy, the higher the probability of it belonging to a true pair. This can be one reason why galaxy interactions produce an increase in star formation (see e.g. \citealp{Patton2013,Pan2018}), because these galaxies have more gas. Although an SFR parameter is included within the \textsc{Caesar} properties, we do not see an effect on it in our results in \Sec{sec:rf_theoretical}, which could be because the interacting galaxies have not had time yet to form stars from this gas. The fact that true pairs tend to have different gas contents in their galaxies (that is, a low $M_\mathrm{H_2,2}/M_\mathrm{H_2,1}$), can be interpreted as a stripping effect of the gas from one of the galaxies by the other. We also want to note that, although for this section we decided to study the molecular gas content of the galaxies, the same general situation is seen when studying the HI fraction instead. 

To summarize this section, we have first seen how all the AHF and \textsc{Caesar} properties available in the simulations can be used to classify observed pairs into `true' or `false'. For that we have used a random forest algorithm that informs us about the importance of each property. This way, we have learned the role that the different properties play when discerning if an observed pair is gravitationally bound or not. We have presented the results here, which indicate that the main properties of interest when doing this kind of classification are the $M_{200}$ and $R_{200}$ ratio of the two galaxies in the pair, together with the spin parameter, the shape of the galaxies, their stellar-to-halo mass ratio and their gas content. Future studies of galaxy pairs should aim at investigating these properties (or other properties that trace them), in order to distinguish bound pairs and understand their peculiarities. In a more observational approach, we have selected a reduced number of properties available to observers, and seen how they alone can be used to classify pairs with a performance that clearly improves that obtained with the traditional observers' method. We thus suggest machine learning as a tool to classify galaxy pairs in future studies, as depending on the information available, this can strongly improve the quality of the obtained samples. We have finally studied how these selected properties affect the distinction between true and false pairs, and tried to understand the physical situation behind this.

\section{Conclusions} \label{sec:conclusions}

In this work, we studied close pairs of galaxies and how we can distinguish if they are gravitationally bound or not. This way, apart from differentiating physical pairs from spurious ones (i.e. created by projection effects), we also distinguish them from flybys (i.e. galaxies that are close but just passing by, not bound to each other). Identifying these bound pairs can be particularly interesting because they are the ones that are most likely to merge in the future. In order to do this, we work with a set of numerical simulations of galaxy clusters and their surroundings, so that we can study galaxy pairs in a cosmological environment. 
We find pairs in the sky following an observer's approach and then we investigate if they are gravitationally bound using the full information from the simulations. We further apply a machine learning algorithm to classify the pairs with improved performance and to understand which features determine this classification.

The simulations used in this work are provided by \textsc{The Three Hundred} project, and consist of a set of 324 numerically modelled spherical regions centred on the most massive clusters found in a prior DM-only cosmological simulation. These 324 regions of radius 15 $\hMpc$ have been re-simulated including full hydrodynamics. For each of them, we limit our study to the region within $5R_{200}$ of the cluster halo centre and select all the objects with $M_\mathrm{*} > 10^{9.5}$ $\Msun$ and $M_\mathrm{200} < 10^{13}$ $\Msun$. 
We project their 3D coordinates (positions and velocities) into the XY plane, thus creating `mock observations'. We then applied the same techniques used by observers to find close pairs of galaxies, based on setting a maximum separation in the sky, $r_\mathrm{sep}$ and a maximum separation in velocity along the line-of-sight, $v_\mathrm{sep}$, for two galaxies to be considered as close. Based on the literature, we used three different values for each of these parameters (20, 50 and 100 $\hkpc$ for $r_\mathrm{sep}$ and 300, 500 and 1000 km/s for $v_\mathrm{sep}$) and combined them. We also allowed for a galaxy to be part of two different pairs. Since we are interested in gravitationally bound galaxies, we kept these as two separate pairs rather than combining them into a group.

After finding the galaxy pairs in the 324 different regions, we compared them against the theoretical study done by \citet{Haggar2021} with the same data, where they find galaxies that are bound to each other based on the criterion in \Eq{eq:roan}. We classified all our observed pairs into `true' (gravitationally bound) or `false' (not bound) and, using \citet{Haggar2021} as the ground truth, in \Sec{sec:stats} we analysed the performance of the observational methods in finding bound pairs. We computed the purity (fraction of observed pairs classified as true, \Fig{fig:purity}) and the pair-completeness (fraction of pairs from \citealp{Haggar2021} that are also found in this work, \Fig{fig:completeness}). We saw that for the most restrictive definitions of proximity ($r_\mathrm{sep}=20$ $\hkpc$), purity can be as high as 82 per cent, but at the cost of missing a significant fraction of the pairs (completeness of 40 per cent). When relaxing the criteria ($r_\mathrm{sep}=100$ $\hkpc$), completeness can be increased to more than 80 per cent but with a purity of around 40 per cent.

In order to improve this classification, and find a method that maximises both purity and completeness, we trained a machine learning algorithm to classify observed galaxy pairs based on their properties. ML has the advantage that it is unbiased and it can handle large amounts of information. 
We used this with all AHF and \textsc{Caesar} properties, which are two different codes to identify halos and galaxies within the simulations (cf. \Sec{sec:data}). These were combined to provide two different approaches -- AHF more theoretical and \textsc{Caesar} more observational -- and obtain as much information as possible as well as checking for consistency between them. We gave all these properties (see Tables~\ref{table:allprops_ahf} and~\ref{table:allprops_caesar}) as inputs to a random forest classifier, trained to differentiate `true' from `false' pairs. 
We did this in two ways, working first with the individual galaxies and using their properties directly (\Fig{fig:rf_props_ahf_caesar}); and then working with the pairs, computing, for each property, the ratio between the two member galaxies (\Fig{fig:rf_ratios_ahf_caesar}). After training the RF model with our data, we obtained the relative importance of each feature and thus learned which properties are most important when carrying out this classification task. The results obtained are summarised below:

\begin{itemize}
    \item The most important feature when classifying the pairs is the radius of the subhalos, $R_{200}$, followed by other related quantities such as the mass $M_{200}$ and the escape velocity. The stellar and gas radii given by \textsc{Caesar} are also found to be relevant. All these quantities refer in the end to the size of the galaxy, for which we find that the relation between the two pair members is important.
    \item Another important feature is the spin parameter, both of all particles and stellar only. It is interesting that these parameters are only relevant for individual galaxies, and not their ratios for the pairs, indicating that the relation between the two galaxies' spin parameters is not affected by them being gravitationally bound. Although there is a correlation between spin and halo mass \citep{KnebePower2008}, the spin parameter also depends on the angular momentum of the subhalo and its orientation, suggesting that this result is not only due to this correlation.
    \item The shape of the stellar content also plays an important role, quantified here by the $b_*$ and $c_*$ parameters, which indicate the degree of sphericity of this component. This confirms our previous results in \citet{Contreras2022b}, where we concluded that galaxies in real pairs tend to have different shapes, an effect that we attributed to stripping. A similar situation is found for the stellar-to-halo mass ratio, with the two galaxies showing more different values. This is also highlighted in this work, and explained in \citet{Contreras2022b} as one of the DM haloes being stripped by the other one. The gas content of galaxies is also shown to be relevant, with similar conclusions to those for the shape and stellar-to-halo mass ratio.
    \item In general, we find that the properties of the most massive galaxy are more important than those of the least massive one when classifying the pair as true or false.
    \item Finally, although the results are shown only for $r_\mathrm{sep}=100$ $\hkpc$ and $v_\mathrm{sep}=500$ km/s, all these conclusions hold when repeating the same procedure using different combinations of the $r_\mathrm{sep}$ and $v_\mathrm{sep}$ parameters.
\end{itemize}

Although this gives us relevant information about the properties of gravitationally bound galaxy pairs, it is a very theoretical approach, and is difficult to exploit from an observational side. For this reason, we then made a selection of observable properties that were also highlighted in the previous step. Working only with the stellar spin parameter, stellar shape, gas content, radius enclosing 80 per cent of the stars and mean stellar metallicity, we trained a new set of random forest algorithms to classify the pairs. When evaluating its performance, we saw that the purity and completeness are better than those computed in \Sec{sec:stats}, reaching 82 and 50 per cent respectively for $r_\mathrm{sep}=100$ $\hkpc$ and $v_\mathrm{sep}=500$ km/s when using all the selected properties. We also saw that different combinations of a number of these properties can already provide a good performance. For instance, using only the shape parameter $c_\mathrm{*}$ of the most massive galaxy in the pair already gives more than 70 per cent in purity and more than 30 per cent in completeness. We analysed these same results for different $r_\mathrm{sep}$ - $v_\mathrm{sep}$ combinations in \Tab{table:rf_results}, where we saw that the general trends remain regardless of these thresholds.

We concluded from here that these RF algorithms can be used to classify galaxy pairs as gravitationally bound or not, based only on a few observable properties. Investigating these properties individually, we determined that galaxies in bound pairs are more likely to have a higher spin parameter and gas content and a less spherical shape than galaxies in spurious pairs or flybys. Additionally, the two bound galaxies in the pair are generally quite different in stellar radii, gas content and shape, an effect that can be attributed to the interaction between them.

We further note here that our study has been carried out for galaxy pairs in cluster environments, reaching up to $5R_{200}$ of the main cluster centre (additional massive objects can also be found here). This means that our results are only valid for these high-density regions and they may not hold for randomly selected pairs from wide surveys, for instance. Regarding this, however, several works have studied galaxy pairs as a function of the environment. Although galaxy interactions are present everywhere in the Universe, their observational manifestations are found to depend on the environment of the galaxies. In general, the results show that the effects of interactions appear to be largest in the lower density environments \citep{Ellison2010,Tonnesen2012,Kampczyk2013,Das2021}. Studies in high-density environments like ours can thus help to clarify this situation and investigate to what extent this is affecting galaxy evolution in clusters. 

Finally, we also want to highlight the importance of creating galaxy pair samples that are as pure and complete as possible, so that they do not introduce any bias that can alter the derived results. Studies like \citet{Bottrell2022} focus on the selection of a pure and complete sample, but for the identification of galaxy merger remnants. Our proposed method for identifying bound galaxy pairs increases significantly the performance of current observational techniques. For this method, we adopted a general approach, so as to not have a biased perspective, but for further work, it would be interesting to focus even more on the properties we highlight here and study in more depth how they affect and are affected by galaxy interactions and mergers. Another task we consider for future work is extending the study to higher redshifts so that we can see if our results hold in time, and analyse what is the fate of gravitationally bound galaxies.

\section*{Acknowledgements}
\addcontentsline{toc}{section}{Acknowledgements}

This work has been made possible by \textsc{The Three Hundred} (\url{https://the300-project.org}) collaboration. The simulations used in this paper have been performed in the MareNostrum Supercomputer at the Barcelona Supercomputing Center, thanks to CPU time granted by the Red Espa\~{n}ola de Supercomputaci\'on. As part of \textsc{The Three Hundred} project, this work has received financial support from the European Union’s Horizon 2020 Research and Innovation programme under the Marie Sklodowskaw-Curie grant agreement number 734374, the LACEGAL project. 

ACS, AK, WC, and GY thank the Ministerio de Ciencia e Innovaci\'{o}n (MICINN) for financial support under research grant PID2021-122603NB-C21. AK further thanks The Who for Quadrophenia. WC is additionally supported by the STFC AGP Grant ST/V000594/1 and the Atracci\'{o}n de Talento Contract no. 2020-T1/TIC-19882 granted by the Comunidad de Madrid in Spain. He further acknowledges the science research grants from the China Manned Space Project with NO. CMS-CSST-2021-A01 and CMS-CSST-2021-B01.

\section*{Data Availability}
The results shown in this work use data from \textsc{The Three Hundred} galaxy clusters sample. These data are available on request following the guidelines of \textsc{The Three Hundred} collaboration, at \url{https://www.the300-project.org}. The data specifically shown in this paper will be shared upon request with the authors.



\clearpage
\bibliographystyle{mnras}
\bibliography{archive}

\appendix

\section{List of properties}  \label{appendix:props}
In \Tab{table:allprops_ahf} we list all the AHF properties used as inputs for the random forest algorithm described in \Sec{sec:results}, together with a brief description of each of them. \Tab{table:allprops_caesar} shows the corresponding list for the \textsc{Caesar} properties.

\begin{table*}
\centering
\caption{List of properties of the galaxies given by the Amiga Halo Finder (AHF) and selected for this work. These properties, together with those in \Tab{table:allprops_caesar} were used as input for a random forest algorithm to classify galaxy pairs into gravitationally bound or not (see \Sec{sec:rf_theoretical}). The index and `name' columns refer to the label given to each property in Figs~\ref{fig:rf_props_ahf_caesar} and~\ref{fig:rf_ratios_ahf_caesar}, while throughout the text they may also be addressed by their `symbol'.}
\begin{tabular}{rlccl}
\hline
 & Name & Symbol & Units & Description \\ \hline
1 & \texttt{M{\_}200}       &    $\mathrm{M_{200}}$    &  $\hMsun$     &   Mass enclosed in the radius at overdensity 200         \\
2 & \texttt{R{\_}200}       &   $\mathrm{R_{200}}$     &   $\hkpc$   &     Radius at overdensity 200        \\
3 & \texttt{Rmax}        &   $\mathrm{R_{max}}$     &   $\hkpc$    &     Position of rotation curve maximum        \\
4 & \texttt{r2}          &   $\mathrm{r^{2}}$     &  $\hkpc$    &     Position where $\rho r^2$ peaks, where $\rho$ is the density    \\
5 & \texttt{mbp{\_}offset}  &        &   $\hkpc$    &    Offset between most bound particle and halo centre         \\
6 & \texttt{com{\_}offset}  &        &   $\hkpc$    &    Offset between centre-of-mass and halo centre         \\
7 & \texttt{Vmax}     &   $\mathrm{V_{max}}$     &   km/s   &  Maximum of rotation curve        \\
8 & \texttt{v{\_}esc}     &   $\mathrm{v_{esc}}$ &   km/s   &  Escape velocity at $\mathrm{R_{200}}$  \\
9 & \texttt{sigV}     &    $\sigma$    &  km/s     &     3D velocity dispersion for all the particles inside the halo        \\
10 & \texttt{lambda}     &    $\lambda$    &   --    &    Spin parameter (\citealp{Bullock2001} definition)  \\
11 & \texttt{lambdaE}     &   $\lambda_\mathrm{E}$     &   --    &     Classical spin parameter (\citealp{Peebles1969} definition)       \\ 
12: 14 & \texttt{Lx,y,z}      &   $\mathrm{L_{x}},\mathrm{L_{y}},\mathrm{L_{z}}$  &  --  &    3 components of the angular momentum vector (with $\vert L \vert = 1$)     \\
15 & \texttt{b}       &   b   &   b/a   &  Ratios of the second major (and minor) to the major axis of the moment of            \\
16 & \texttt{c}       &   c   &   c/a   &    \, inertia tensor (a value equal to 1 indicates perfect sphericity)         \\
17 & \texttt{Eax}     &   $\mathrm{E_{a,x}}$     &       &      Largest axis of moment of inertia tensor   (with $\vert E \vert = 1$)    \\
18 & \texttt{Eay}     &   $\mathrm{E_{a,y}}$     &       &             \\
19 & \texttt{Eaz}     &   $\mathrm{E_{a,z}}$     &       &             \\
20: 22 & \texttt{Ebi}     &  $\mathrm{E_{b,i}}$      &       &    Second largest axis of moment of inertia tensor (3 components)         \\
23: 25 & \texttt{Eci}     &  $\mathrm{E_{c,i}}$      &       &    Third largest axis of moment of inertia tensor  (3 components)       \\
26 & \texttt{Ekin}     &   $\mathrm{E_{kin}}$     &  $\hMsun$ ${(\mathrm{km/s})}^2$  &    Kinetic energy         \\
27 & \texttt{Epot}     &    $\mathrm{E_{pot}}$    &   $\hMsun$ ${(\mathrm{km/s})}^2$    &     Potential energy        \\
28 & \texttt{SurfP}     &    $\mathrm{P_{s}}$    &   $\hMsun$ ${(\mathrm{km/s})}^2$    &      Surface pressure (\citealp{Shaw2006} definition)       \\
29 & \texttt{M{\_}star}     &   $\mathrm{M_{*}}$     &   $\hMsun$    &      Mass of stellar particles       \\
30: 31 & \texttt{lambda(E){\_}star}  &   $\lambda_{*}$, $\lambda_{E,*}$     &   --    &       Spin parameters for stars      \\
32: 34 & \texttt{Lj{\_}star}     &  $\mathrm{L_{j,star}}$     &     --  &      3 components of the stellar angular momentum vector       \\
35: 36 & \texttt{b{\_}star}, \texttt{c{\_}star}     &    $\mathrm{b}_*$, $\mathrm{c}_*$    &   --    &     b and c parameters for stellar components        \\
37: 45 & \texttt{Eji{\_}star}     &   $\mathrm{E_{j,i,*}}$  &  --  &  3 (i) components of the three (j) axes of the stellar moment of inertia tensor            \\
46 & \texttt{Ekin{\_}star}     &   $\mathrm{E_{kin,*}}$     &   $\hMsun$ ${(\mathrm{km/s})}^2$    &     Kinetic and potential energy for stars        \\
47 & \texttt{Epot{\_}star}     &   $\mathrm{E_{pot,*}}$     &   $\hMsun$ ${(\mathrm{km/s})}^2$    &             \\
48 & \texttt{mean{\_}z{\_}star}  &   $Z_*$     &    $Z_\odot$   &     Mean stellar metallicity        \\
49 & \texttt{SMHM}     &    $\mathrm{M}_* / \mathrm{M_{200}}$     &   --    &   Stellar-to-halo mass ratio          \\ \hline
50 & \texttt{JOHNSON{\_}V}    &    Johnson-V    &   --    &      Absolute magnitudes in the different spectral bands as computed by the      \\
51 & \texttt{JOHNSON{\_}B}    &   Johnson-B     &   --    &     \, stellar population synthesis code \textsc{stardust} \citep{Devriendt99} by          \\
52 & \texttt{JOHNSON{\_}H}    &   Johnson-H     &   --    &     \, considering the contribution of all the individual stellar particles        \\
53 & \texttt{JOHNSON{\_}I}    &   Johnson-I     &   --    &             \\
54 & \texttt{JOHNSON{\_}J}    &   Johnson-J     &   --    &             \\
55 & \texttt{JOHNSON{\_}K}    &   Johnson-K     &   --    &             \\
56 & \texttt{JOHNSON{\_}R}    &   Johnson-R     &   --    &             \\
57 & \texttt{JOHNSON{\_}U}    &   Johnson-U     &   --    &             \\
58 & \texttt{U-B}     &   $U-B$     &   --    &     \, Colour index obtained subtracting the indicated magnitudes: \\
    &   &   &    & \,  Johnson-U $-$ Johnson-B        \\
59 & \texttt{B-V}     &   $B-V$     &   --    &     \, Johnson-B $-$ Johnson-V        \\
60 & \texttt{V-I}     &   $V-I$     &   --    &     \, Johnson-V $-$ Johnson-I        
\end{tabular}
\label{table:allprops_ahf}
\end{table*}

\begin{table*}
\centering
\caption{List of properties of the galaxies given by \textsc{Caesar} galaxy finder and selected for this work. These properties, together with those in \Tab{table:allprops_ahf} were used as input for a random forest algorithm to classify galaxy pairs into gravitationally bound or not (see \Sec{sec:rf_theoretical}). The index and `name' columns refer to the label given to each property in Figs~\ref{fig:rf_props_ahf_caesar} and~\ref{fig:rf_ratios_ahf_caesar}, while throughout the text they may also be addressed by their `symbol'. By default, length units are comoving kpc.}
\begin{tabular}{rlccl}
\hline
 & Name & Symbol & Units & Description \\ \hline
61 & \texttt{masses: H2}   &  $\mathrm{M_{H_2}}$   &  $\Msun$     & $\mathrm{H}_2$ and HI masses come from assigning all the gas in the \\
62 & \texttt{masses: HI}              &   $\mathrm{M_{HI}}$     &  $\Msun / \mathrm{M_{b}}$    &    \, halo to its most bound galaxy within the halo,         \\ 
63 & \texttt{masses: baryon}          &  $\mathrm{M_{b}}$      &  $\Msun$   &     \, \texttt{baryon} includes both stellar and gas particles,       \\ 
64 & \texttt{masses: gas}             &   $\mathrm{M_{gas}}$     &  $\Msun$   &              \\
65 & \texttt{masses: gas{\_}stellar{\_}half{\_}mass}   & $\mathrm{M_{gas,shm}}$  & $\Msun$  &  \, \texttt{{\_}stellar{\_}half{\_}mass} ($\mathrm{shm}$) denotes the radii enclosing 50 per  \\ 
66 & \texttt{masses: star{\_}stellar{\_}half{\_}mass}  & $\mathrm{M_{*,shm}}$  & $\Msun$  &  \, \, \,  cent of the stellar mass  \\ 
67 & \texttt{masses: stellar}         &  $\mathrm{M_{*}}$      & $\Msun$      &             \\
68 & \texttt{radii: baryon{\_}half{\_}mass}     &   $\mathrm{R_{b,50}}$     &    kpc   &     Radius enclosing 50 per cent of baryons        \\
69 & \texttt{radii: baryon{\_}r20}     &   $\mathrm{R_{b,20}}$     &   kpc    &    Radius enclosing 20 per cent of baryons         \\
70 & \texttt{radii: baryon{\_}r80}     &   $\mathrm{R_{b,80}}$     &   kpc    &    Radius enclosing 80 per cent of baryons         \\
71: 73 & \texttt{radii: gas{\_}XX}     &  $\mathrm{R_{gas,XX}}$      &   kpc    &     Same for gas and stars (the galaxy center of mass from which     \\
74: 76  & \texttt{radii: stellar{\_}XX}     &   $\mathrm{R_{*,XX}}$     &   kpc    &   \,  the radii are found is recomputed for each type)            \\
77 & \texttt{rotation: baryon{\_}ALPHA}     &   $\alpha_\mathrm{b}$     &   --    &   $\alpha$ and $\beta$ are the rotation angles required to rotate the galaxy        \\
78 & \texttt{rotation: baryon{\_}BETA}     &   $\beta_\mathrm{b}$     &   --    &     \, to align with the angular momentum        \\
79 & \texttt{rotation: baryon{\_}BoverT}     &   $\mathrm{B/T_{b}}$     &   --    &    Bulge-to-total mass ratio, where the bulge mass is \\
    &   &   &   & \, defined kinematically as twice the counter-rotating mass         \\
80 & \texttt{rotation: baryon{\_}L{\_}coord0}     &  $\mathrm{L_{0,b}}$      &  $\Msun \cdot $kpc$\cdot$km/s     &    3 components of the angular momentum vector of the        \\
81 & \texttt{rotation: baryon{\_}L{\_}coord1}     &  $\mathrm{L_{1,b}}$      &  $\Msun \cdot $kpc$\cdot$km/s     &   \, galaxy          \\
82 & \texttt{rotation: baryon{\_}L{\_}coord2}     &  $\mathrm{L_{2,b}}$      &  $\Msun \cdot $kpc$\cdot$km/s     &             \\
83 & \texttt{rotation: baryon{\_}kappa{\_}rot}     &  $\kappa_\mathrm{rot,b}$      &   --    &   Fraction of kinetic energy in rotation \citep{Sales2012}          \\
84: 90 & \texttt{rotation: gas{\_}XX}     &        &       &     Same for gas and stars        \\
91: 97 & \texttt{rotation: stellar{\_}XX}     &        &       &             \\
98 & \texttt{v{\_}disps: baryon}     &   $\sigma_\mathrm{b}$     &   km/s    &    Mass-weighted velocity dispersions for each particle type,           \\
99 & \texttt{v{\_}disps: gas}     &   $\sigma_\mathrm{gas}$     &   km/s    &    \, computed around the centre of mass velocity (recomputed         \\
100 & \texttt{v{\_}disps: gas{\_}stellar{\_}half{\_}mass}     &    $\sigma_\mathrm{gas,shm}$    &   km/s    &    \, for each type)     \\
101 & \texttt{v{\_}disps: star{\_}stellar{\_}half{\_}mass}     &   $\sigma_\mathrm{*,shm}$     &   km/s    &         \\
102 & \texttt{v{\_}disps: stellar}     &  $\sigma_\mathrm{*}$      &  km/s     &             \\
103 & \texttt{absmag: sdss{\_}g}     &   sdss-$g$     &   --    &   Absolute magnitudes for the indicated photometric band          \\
104 & \texttt{absmag: sdss{\_}i}     &   sdss-$i$     &   --  &             \\
105 & \texttt{absmag: sdss{\_}r}     &   sdss-$r$     &   --    &             \\
106 & \texttt{absmag: sdss{\_}u}     &   sdss-$u$     &   --    &             \\
107 & \texttt{absmag: sdss{\_}z}     &   sdss-$z$     &   --    &             \\
108 & \texttt{colours: g-r}     &   $g-r$     &   --    &     Colour indices obtained by subtracting the indicated         \\
109 & \texttt{colours: u-r}     &   $u-r$     &   --    &  \,   absolute magnitudes from above        \\
110 & \texttt{colours: r-i}     &   $r-i$     &   --    &             \\
111 & \texttt{ages: mass{\_}weighted}     &    age    &   Gyr   &    Mean stellar age, weighted by mass         \\
112 & \texttt{metallicities: sfr{\_}weighted}     &    $Z_\mathrm{gas}$    &   --    &    Gas-phase metallicity, weighted by SFR, in total metal         \\
 &  &    &      &    \, mass fractions (not solar-scaled)          \\
113 & \texttt{metallicities: stellar}     &   $Z_\mathrm{*}$     &   --    &     Stellar metallicity, mass weighted        \\
114 & \texttt{sfr}     &   SFR     &   $\Msun$/yr    &    Instantaneous star formation rate, from summing SFR        \\
    &    &   &   & \, in gas particles \\
115 & \texttt{sfr{\_}100}     &     $\mathrm{SFR_{100}}$   &   $\Msun$/yr    &   SFR averaged over last 100 Myr, from star particles \\
    &   &   &    & \,  formed in that time  
\end{tabular}
\label{table:allprops_caesar}
\end{table*}

\section{Validation of machine learning models}  \label{appendix:ml_validation}

\begin{figure}
   \includegraphics[width=8cm]{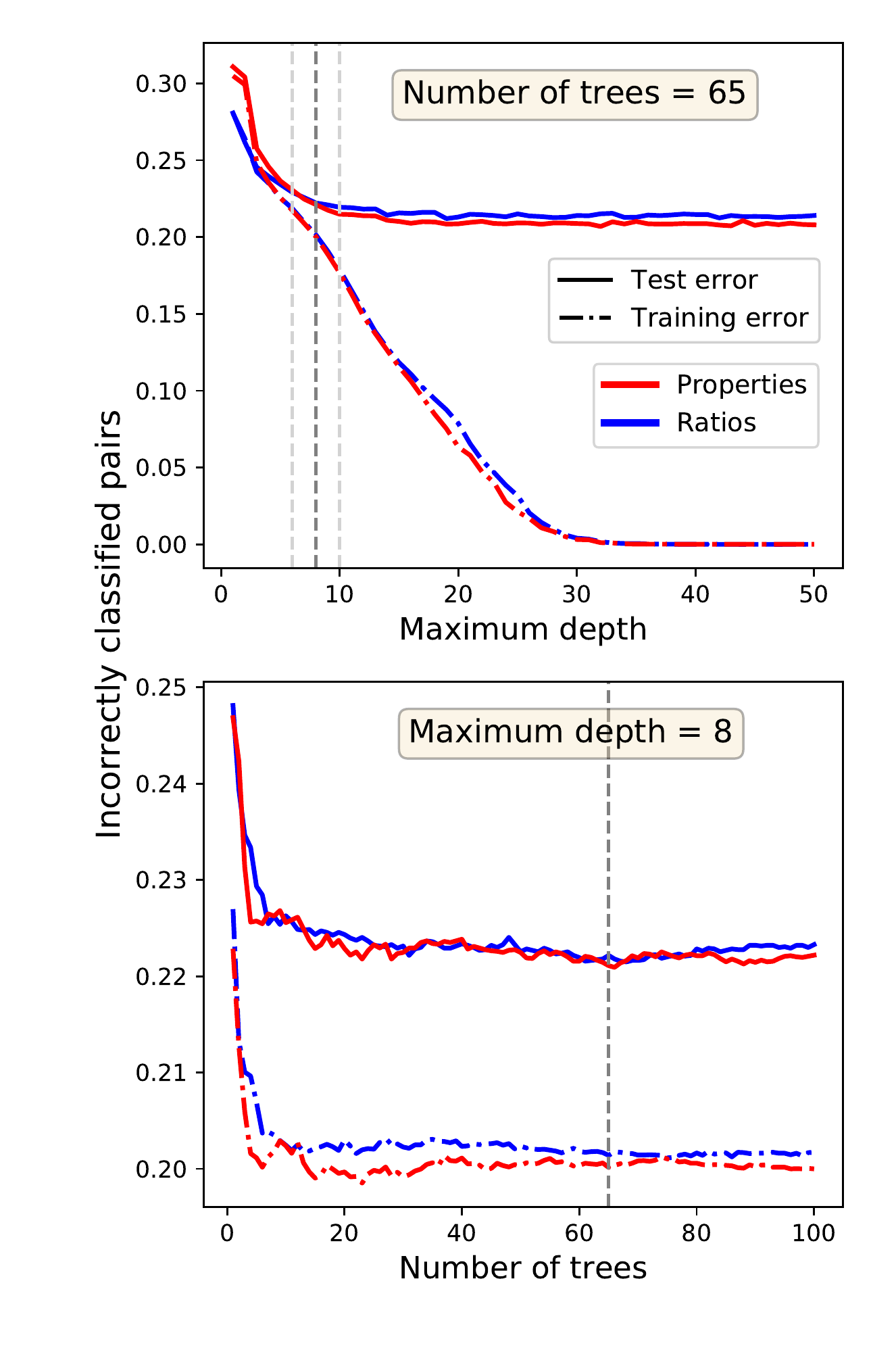}
   \vspace{-0.3cm}\caption{\textbf{Top,} training (dash-dotted) and test (solid) errors of the random forest models presented in \Sec{sec:rf_theoretical} (red for individual properties of galaxies -- \Fig{fig:rf_props_ahf_caesar}, and blue for the ratios between the two pair members -- \Fig{fig:rf_ratios_ahf_caesar}), as a function of the maximum depth allowed for the decision trees (for a fixed number of trees of 65). Vertical dashed lines indicate values of 6, 8 and 10. A maximum depth of 8 is selected for these models, since it minimizes the test error while keeping the training error similar. \textbf{Bottom,} same but as a function of the number of trees used, for a fixed maximum depth of 8. Note the different scale in the $y$-axis. The vertical dashed line indicates the value of 65 used throughout the paper.}
\label{fig:ml_validation}
\end{figure}

In this section we show the validation of the machine learning models that we introduced in Sections~\ref{sec:rf_theoretical} and~\ref{sec:rf_observational}. For this we compute the training and test error of each of the models -- that is, the fraction of pairs in the training and test sets, respectively, that are incorrectly classified as either true or false by the model. 
In \Fig{fig:ml_validation} we show these two errors for the two models in \Sec{sec:rf_theoretical}, i.e. one using the properties of individual galaxies (in red), and the one for the ratios of each property between the two pair members (in blue). In this figure we run the random forest classifiers with different parameters and analyse the performance of the models in each of the cases, in order to choose the best one. In the top panel of \Fig{fig:ml_validation} we plot these errors as a function of the maximum depth allowed for the individual trees in the random forest, for a fixed number of trees. The maximum depth of a tree is the largest possible length between the root to a leaf. In the plot we can see that, while both training and test errors are similar for the smaller maximum depths, the training error decreases fast for increasing maximum depth. The behaviour of the errors for the higher maximum depths indicate that our models are overfitted -- or overtrained --, since they are working much better on the training set than on the test set. The high values of the maximum depth allow the decision trees to grow so as to fit perfectly the training set, instead of adapting to an arbitrary test set. To avoid overfitting, we select a maximum depth that reduces the test error as much as possible while keeping a training error that is similar to the test error. We choose a value of 8, indicated in \Fig{fig:ml_validation} as a vertical dashed line, and fix this maximum depth for all the models. 

In the bottom panel of \Fig{fig:ml_validation} we now fix the maximum depth to 8 and plot the training and test errors of the models as a function of the number of decision trees used. In this case we see that both errors converge to a reasonably stable value within $\sim 20$ trees. The number of trees that minimises the test error for both models is 65, and thus we fix this parameter at this value, although we can see in \Fig{fig:ml_validation} that the precise value used for the number of trees will not affect very significantly the results in terms of training and test errors of the models.
We want to highlight that, for the selected values for the maximum depth and the number of trees of the model, the difference between training and test error is below 10 per cent.

\begin{figure}
    \centering
   \includegraphics[width=6cm]{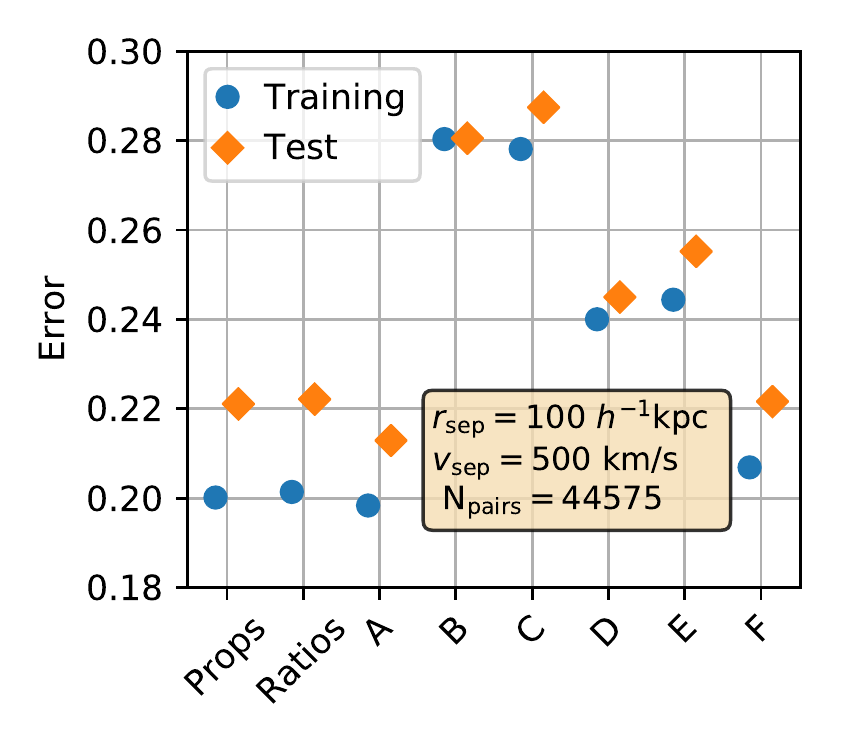}
   \caption{Training (blue dots) and test (orange diamonds) errors of the different random forests presented in this paper, for fixed number of trees of 65 and maximum depth of 8. The first two values in the $x$-axis represent the two theoretical models in \Sec{sec:rf_theoretical}, while the models from A to F are those presented in \Sec{sec:rf_obs_1}, which use only a subselection of properties.}
\label{fig:ml_validation_simple}
\end{figure}

Finally, in order to validate all the machine learning models presented, in \Fig{fig:ml_validation_simple} we show their training and test error for fixed values of number of trees of 65 and maximum depth of 8. The training error is plotted as blue dots, while the test error is depicted by orange diamonds. We also indicate the total number of pairs used in this model, 44575 in this case, together with a reminder of the distance and line-of-sight velocity separation thresholds used, $r_\mathrm{sep} = 100$ $\hkpc$ and $v_\mathrm{sep} = 500$ km/s, respectively. The first two values in the $x$-axis in \Fig{fig:ml_validation_simple} represent the two theoretical models in \Sec{sec:rf_theoretical} (same as \Fig{fig:ml_validation} but for fixed number of trees and maximum depth). The following ticks, A to F, indicate the models presented in \Sec{sec:rf_obs_1}, constructed with only a subselection of the properties found to be most relevant as well as available in observations. In \Fig{fig:ml_validation_simple} we can see that, although the test error is always higher than the training error, they are always within a difference below 10 per cent, showing that no overfitting is taking place. This is especially true for the A-F models, where the difference between the two errors is even smaller.

\begin{table}
\centering
\caption{For each combination of projected distance, $r_\mathrm{sep}$, and line-of-sight velocity separation, $v_\mathrm{sep}$, thresholds, parameters of the random forest models used to classify pairs: total number of pairs (i.e. size of the sample), number of decision trees used, and maximum depth allowed for each of them.}
\begin{tabular}{r|c|c|c|c}
\hline
 {$\bm{r_\mathrm{sep}}$} &     $\bm{v_\mathrm{sep}}$    & \multirow{2}{*}{$\bm{N_\mathrm{pairs}}$}   &\multirow{2}{*}{$\bm{n_\mathrm{trees}}$}  &\multirow{2}{*}{\textbf{max$\_$depth}} \\
{{($\hkpc$)}} & {(km/s)}  &    &  &   \\ \hline \hline
\textbf{100}  & \textbf{500}   & 44575   & 65     & 8            \\ \hline
\textbf{20}  & \textbf{300}    & 5009    & 65      & 6             \\ 
             & \textbf{500}    & 5720    & 55      & 6             \\ 
             & \textbf{1000}   & 6945    & 55      & 6             \\ \hline 
\textbf{50}  & \textbf{300}    & 12182   & 60      & 6             \\ 
             & \textbf{500}    & 15815   & 65      & 7             \\
             & \textbf{1000}   & 22795   & 65      & 8             \\ \hline 
\textbf{100} & \textbf{300}    & 31007   & 65      & 8             \\ 
             & \textbf{1000}   & 71608   & 65      & 8       \\ \hline
\end{tabular}
\label{table:ml_validation}
\end{table}

The process shown here in Figs~\ref{fig:ml_validation} and~\ref{fig:ml_validation_simple} was repeated for all the different combinations of thresholds in distance, $r_\mathrm{sep}$, and line-of-sight velocity separation, $v_\mathrm{sep}$, between pair members. This means we find the optimal values for the maximum depth and number of trees of the models, so that the training and test error are within 10 per cent of one another. In \Tab{table:ml_validation} we summarise the properties of the different models -- that is, we include for each $r_\mathrm{sep}$ - $v_\mathrm{sep}$ combination the total number of pairs used (i.e. the size of the sample), the number of trees used, and the maximum depth allowed for them. The results in \Tab{table:rf_results} regarding the purity and completeness of this models were obtained with the parameters shown here.


\bsp	
\label{lastpage}
\end{document}